\documentclass[onecolumn,amsmath,amssymb,11pt]{revtex4}

\def\gsim{ \lower .75ex \hbox{$\sim$} \llap{\raise .27ex \hbox{$>$}} }

\def\lsim{ \lower .75ex \hbox{$\sim$} \llap{\raise .27ex \hbox{$<$}} }

\usepackage{graphicx}
\usepackage{dcolumn}
\usepackage{bm}
\input epsf
\def\gsim{ \lower .75ex \hbox{$\sim$} \llap{\raise .27ex \hbox{$>$}} }
\def\lsim{ \lower .75ex \hbox{$\sim$} \llap{\raise .27ex \hbox{$<$}} }

\usepackage{graphicx}
\usepackage{dcolumn}
\usepackage{bm}

\newcommand{\nn}{\nonumber}
\newcommand{\be}{\begin{equation}}
\newcommand{\ee}{\end{equation}}
\newcommand{\bea}{\begin{eqnarray}}
\newcommand{\eea}{\end{eqnarray}}
\renewcommand{\d}{\mathrm{d}}
\def\a{\alpha}

\def\d{\delta}

\def\s{\sigma}
\def\sb{\bar{\sigma}}

\def\l{\lambda}
\def\m{\mu}
\def\n{\nu}

\def\p{\phi}
\def\pb{\bar\psi}

\def\th{\theta}
\def\tb{\bar{\theta}}

\def\s{\sigma}

\def\t{\tau}

\def\pt{\partial}

\def\ad{\dot\alpha}

\def\Db{\bar{D}}
\def\As{A^\ast}
\def\Fs{F^\ast}
\def\P{\Phi}
\def\Pd{\Phi^\dag}

\begin{document}

\title{Supersymmetric $P(X,\phi)$ and the Ghost Condensate}

\author{Justin Khoury$^1$, Jean-Luc Lehners$^{2,3}$ and Burt Ovrut$^1$}

\affiliation{$^1$ Department of Physics, University of Pennsylvania, 209 South 33rd Street, Philadelphia, PA 19104-6395, U.S.A.\\$^2$ Perimeter Institute for Theoretical Physics, Waterloo, ON N2L 2Y5 Canada\\ $^3$ Max-Planck-Institute for Gravitational Physics (Albert-Einstein-Institute), 14476 Golm, Germany }

\begin{abstract}
We show how to construct supersymmetric actions for higher-derivative scalar field theories of the form $P(X,\phi)$, where $X\equiv -(\pt\phi)^2/2m^4$, within the context of $d=4,$ ${\cal{N}}=1$ supersymmetry. This construction is of general use, and is applied to write supersymmetric versions of the Dirac-Born-Infeld action. Our principal application of this formalism is to construct the supersymmetric extension of the ghost condensate. This allows us to study the interplay between supersymmetry, time-dependent backgrounds and violations of the null energy condition.
\end{abstract}

\maketitle

\section{Introduction}

Generic compactifications of string theory lead to a number of scalar fields, or moduli, in the resulting effective 4-dimensional theory. The moduli describe the compactification geometry, gauge connections and the positions of branes in the internal space. Although at late times one requires these fields to be fixed at a minimum of their potential,
it is generally believed that in the very early universe, close to the time of the big bang or before, some of these moduli might have played an important dynamical role.
The effective actions for these scalar fields contain higher-derivative kinetic terms, which can become significant at larger  energy while remaining within the regime of validity of the theory.

One example is Dirac-Born-Infeld (DBI) inflation~\cite{Silverstein:2003hf,Alishahiha:2004eh}, where the scalar field describing the position of a brane plays the role of the inflaton field and drives a phase of accelerated expansion of the universe. In this scenario, interesting new dynamics arises precisely because of the higher-derivative kinetic terms. Due to non-linearly realized symmetries, the effective DBI theory remains valid for large brane velocity provided the acceleration is small.
The higher-derivative form leads to distinct predictions for cosmological observables -- in this case, equilateral-type non-gaussianity in the primordial density fluctuations~\cite{Alishahiha:2004eh,Babich:2004gb}.

Higher-derivative scalar field actions also play a central role in ekpyrotic theories~\cite{Khoury:2001wf,Donagi:2001fs,Khoury:2001bz,Khoury:2001zk,Craps:2003ai,Khoury:2003vb,Khoury:2003rt,Khoury:2004xi,Lehners:2007ac,Buchbinder:2007ad,Buchbinder:2007tw,Creminelli:2007aq,Koyama:2007ag,Koyama:2007mg} and other bouncing cosmologies~\cite{Gasperini:1992em,Gasperini:2002bn,Finelli:2001sr,Creminelli:2010ba}, in which the big bang is preceded by an era of slow contraction. See~\cite{Lehners:2008vx} for a review. Ekpyrotic ideas were originally presented  within the context of heterotic M-theory~\cite{Lukas:1997fg,Lukas:1998yy,Lukas:1998tt,Lukas:1998qs,Lukas:1998ew,Lukas:1999yn,Braun:2005nv,Braun:2005ux} and heterotic M-theory cosmology~\cite{Lukas:1996zq,Lukas:1996ee,Lukas:1996iq,Lukas:1997yc,Brandle:2000qp}, with the geometric and five-brane moduli playing the essential role~\cite{Buchbinder:2002ic,Buchbinder:2002ji,Lima:2001jc,Lima:2001nh,Donagi:1999jp,Ovrut:2000qi}. In the pre-big bang epoch, quantum fluctuations in the scalar fields lead to a nearly scale invariant spectrum of density perturbations with significant non-gaussianities~\cite{Creminelli:2007aq,Buchbinder:2007at,Koyama:2007if,Lehners:2007wc,Lehners:2008my,Mizuno:2008zza,Lehners:2009qu,Lehners:2009ja,Khoury:2009my,Khoury:2010gw}, as reviewed in~\cite{Lehners:2010fy}. A key prediction is that the primordial gravitational wave spectrum has a strong blue tilt and, hence, negligible amplitude on large scales~\cite{Khoury:2001wf,Boyle:2003km,Baumann:2007zm}.

Central to the viability of bounce scenarios is a smooth cosmological transition from contraction to expansion. This requires violating one of the most important energy conditions in general relativity, the Null Energy Condition (NEC). For this to occur, the cosmology must pass through an epoch in which the sum of the effective energy density and pressure is negative, $\rho + p < 0$. Unfortunately, it was shown in~\cite{Dubovsky:2005xd} that if the ``usual'' scalar contributions to the stress-energy -- that is, theories possessing {\it two}-derivative kinetic energy with
any number of scalar fields, arbitrary relativistic fluids and solids -- violate the NEC, then they necessarily contain  ghosts or gradient instabilities.
However, it was realized that certain higher-derivative theories, specifically those with second-order equations of motion, provide a loophole to this theorem~\cite{Dubovsky:2005xd,Creminelli:2006xe,Nicolis:2009qm,Creminelli:2010ba}.
The first example of this was the {\it ghost condensate}~\cite{ArkaniHamed:2003uy,ArkaniHamed:2005gu}.
This allows for ghost-free NEC violations leading to a smooth bounce~\cite{Creminelli:2006xe}. A second example is the recently proposed Galileon scalar field theories~\cite{Nicolis:2008in,Nicolis:2009qm}.
Whether or not these theories can be realized in string theory has yet to be determined~\cite{Dubovsky:2006vk,Adams:2006sv}. But, from a purely low-energy perspective, they offer consistent ghost-free field theories which achieve NEC violation and, hence,  bouncing cosmology. New Ekpyrotic theory, in particular, successfully merges an ekpyrotic contracting phase with a ghost-condensate phase -- leading to the big bang and the present epoch of expansion~\cite{Buchbinder:2007ad,Buchbinder:2007tw}.

In many cases of interest, the 4-dimensional actions ought to be {\it supersymmetric}. There are several reasons for this. The first is that the configurations in which some amount of supersymmetry is preserved are also those which we trust the most, as the residual supersymmetry protects the action from large and difficult to calculate corrections. Secondly, one expects these  theories to arise from compactifications of superstrings to $4$-dimensions. Such vacua often retain some degree of supersymmetry. From the point of view of phenomenology, the most interesting case to consider is that of minimal ${\cal N}=1$ supersymmetry and, hence, we will restrict our attention to that case. In this paper, we show how to construct supersymmetric extensions of scalar field theories whose kinetic terms are of the form $P(X,\phi)$, where $P$ is an arbitrary function that is analytic in $X\equiv -(\pt\phi)^2/2m^4$ and $\p$ around zero. For example, the above-mentioned DBI theory is of this type
\footnote{Maximally supersymmetric brane actions have been considered extensively in the literature, starting with \cite{Aganagic:1996pe,Aganagic:1996nn,Cederwall:1996ri,Bergshoeff:1996tu}. However, as far as we are aware of, theories of the DBI and more generally $P(X,\phi)$ type have not been constructed in minimal ${\cal N}=1,$ $d=4$ supersymmetry, which is the framework most relevant for phenomenology. For related work in $d=3$ spacetime dimensions, see \cite{Bazeia:2009db}.}.
Hence, our results offer an ${\cal N}=1$ supersymmetric framework in which to study DBI inflation.

Ghost condensate actions are also of this form. The supersymmetrized ghost condensate theory presented here offers a concrete setting in which to explore the relationship between supersymmetry, time-dependent backgrounds and NEC violation. As elaborated below, we find that, not unexpectedly, the marriage of supersymmetry and NEC violation is not one without friction. Indeed, aside from $\phi$, one now has to worry about the stability of the other component fields -- the second real scalar field $\chi$, the spinor $\psi_{\alpha}$ and the auxiliary field $F$ -- required by supersymmetry. (In this paper, we will always take $\phi$ to be one scalar component of a {\it chiral} superfield.) For ``pure" $P(X)$ ghost condensate theory, we find that, as usual, the spatial gradient of $\phi$ perturbations vanishes at the ghost condensate point and becomes negative in the NEC violating region. Nevertheless, the dispersion relation for $\phi$ can be stabilized by including supersymmetric higher-derivative terms that leave the action for the other components fields unchanged at quadratic order. The fluctuations in the second  scalar field $\chi$, however, come out entirely wrong -- the time-derivative piece vanishes, whereas the spatial gradient has the wrong sign. Once again, however, we find supersymmetric corrections to the action that make the $\chi$
fluctuations -- both temporal and spatial --  stable. Finally, the time-derivative kinetic term for the fermions is healthy, but their gradient term is of the wrong sign.
But in this case, we are unable to find a simple modification to the action with the superfield expressions we have analyzed. Thus, within the context of supersymmetric $P(X)$ theories
and their simple generalizations, the fermion spatial gradient term has the wrong sign, though the nature and physical implications of this potential instability remains unclear to us.
It is worth advertising that in the supersymmetric extensions of more general higher-derivative scalar theories, the fluctuations in all component fields -- including spinors -- {\it can} have correct sign temporal and gradient kinetic terms around NEC-violating solutions. This is the case, for example, in Galileon theories. We will present  these results, including a complete supersymmetrization of Galileon theories,
in a forthcoming publication.

The paper is organized as follows. In Sec.~\ref{susyP}, we describe the steps in supersymmetrizing $P(X,\phi)$ theories, starting from the simplest $X$ and $X^2$ terms (Sec.~\ref{XX2}), generalizing to $X^n$ (Sec.~\ref{Xn}) and culminating in the general $P(X,\phi)$ case (Sec.~\ref{Pfull}). In Sec.~\ref{egDBI} we apply these results to construct the supersymmetric version of the DBI action, for arbitrary warp factor. We then turn our attention to supersymmetric ghost condensation (Sec.~\ref{sectionGhostCondensate}), including a complete analysis of perturbations around the ghost condensate solution. We recap the key results and outlook for future directions in Sec.~\ref{recap}.

\section{Supersymmetric $P(X,\phi)$}
\label{susyP}

We will work in superspace, as this approach makes the invariance of our actions under supersymmetry transformations automatic. Our strategy is to first re-write the non-supersymmetric Lagrangian function $P(X,\phi)$ as a series in powers of $X$ with field-dependent coefficients:
\be
P(X,\p) = \sum_{n\in \mathbb{N}^\ast} a_n(\p)X^n\,,\ee where \be X\equiv -\frac{1}{2m^4}(\pt\p)^2 = \frac{1}{2m^4}(\dot\p^2-\p^{,i}\p_{,i})\,.
\ee
Here $m$ is a mass scale introduced so as to render $X$ dimensionless. For most of the paper, we will set this mass equal to $1$, in order to minimize the amount of clutter in our formulae and because it is straightforward to reinstate the $m$-dependence if required. Space-time indices are denoted by $\mu,\nu,...,$ spatial indices by $i,j,...$ and time derivatives by overdots. Our goal is to construct the superspace expressions for the individual terms $a_n X^n,$ using chiral and anti-chiral superfields $\P$ and $\Pd$ respectively, as well as the superspace derivatives $D$ and $\Db$ (our notation and conventions are those of Wess and Bagger \cite{Wess:1992cp}). More precisely, the chiral superfield $\P$ consists of a complex scalar $A \equiv (\p + i\chi)/\sqrt{2},$ with $\p$ and $\chi$ real scalars, an auxiliary field $F$ and a Weyl spinor $\psi_\a$. We will be interested in superspace Lagrangians which reduce to the form $a_n X^n$ when $\chi=F=\psi_{\alpha} = 0.$ In fact, we first analyze the case where the functions $a_n$ are constant. It will then be straightforward to extend our formulae to include the case of field-dependent coefficients.

Before embarking on this program, we should clarify one point. We will construct theories invariant under {\it global} supersymmetry, even though we have in mind applications to cosmology. This is because, as we will see,
the salient features of the supersymmetric extension are already contained in the globally supersymmetric theories. However, since we have in mind eventual coupling to gravity, the overall sign of the actions considered here is meaningful and will, therefore, be chosen appropriately. This will be particularly important when we look at the ghost condensate model.

\subsection{$X$ and $X^2$}
\label{XX2}

In order to set our notation, let us briefly review the standard construction of the supersymmetric extension of $X$. A chiral superfield $\P$ has a component expansion in superspace that reads
\be
\P = A + i\th\s^\mu\tb A_{,\mu} + \frac{1}{4}\th\th\tb\tb\Box A + \th\th F + \sqrt{2} \th\psi -\frac{i}{\sqrt{2}}\th\th \psi_{,\mu}\s^\mu \tb\,,
\ee
with the complex scalar $A(x),$ the auxiliary field $F(x)$ and the spinor $\psi_\a(x)$ being functions of the ordinary space-time coordinates $x^\mu.$ Spinor indices which we do not write out explicitly are understood to be summed according to the convention $\psi \th = \psi^\a \th_\a$ and $\pb \tb = \bar\psi_{\ad} \bar\th^{\ad}.$ The  supersymmetric action is given by
\be
\int {\rm d}^4 x {\rm d}^4\th\, \Pd \P = \int {\rm d}^4 x\, \Pd\P \mid_{\th\th\tb\tb}  \,,
\label{rain1}
\ee
where
\bea
\Pd\P \mid_{\th\th\tb\tb}&=&  \frac{1}{4}\As \Box A + \frac{1}{4} A \Box \As -\frac{1}{2}\pt A \cdot \pt \As +\Fs F -\frac{i}{2} \bar\psi \sb^\mu \psi_{,\mu} + \frac{i}{2} \bar\psi_{,\mu} \sb^\mu \psi \nn \\ &=& -\pt A \cdot \pt \As + \Fs F + \frac{i}{2} (\psi_{,\m}\s^\m \bar\psi - \psi \s^\m \bar\psi_{,\m}) \ .
\label{rain2}
\eea
To obtain the last line we have used integration by parts, as well as an identity involving the $\s$ matrices \cite{Wess:1992cp}. After rewriting the complex scalar $A$ in terms of the real fields $\p$ and $\chi$ as \be A \equiv \frac{1}{\sqrt{2}}(\p + i \chi)\,,\ee we can see that the above action contains $X = -\frac{1}{2}(\pt\p)^2$; becoming identical to it when we set $\chi=F=\psi_{\alpha} = 0.$

We now want to find a supersymmetric version of $X^2.$ As it turns out, this is the most crucial part of our calculation. As we will see, once one has the supersymmetric version of $X^2$, it is straightforward to use this expression as a building block to construct the supersymmetric extensions of higher powers of $X.$ Above, we saw that the superspace integral of $\Pd\P$ gave us a Lagrangian containing
$(\pt\p)^2.$ We now want to generate kinetic terms involving two more fields and two more space-time derivatives. The supersymmetry algebra states that the anticommutator of the superspace derivatives $D$ and $\Db$ is proportional to the momentum operator, which is a space-time derivative,
\be
\{ D_\a , \Db_{\ad} \} = -2i \s^{\mu}_{\a\ad}\pt_{\mu}\,. \label{susyalgebra}
\ee
Hence, we need to insert two more $D$'s and two more $\Db$'s into our superfield expression, as well as another $\P$ and $\Pd.$ (We will comment on superfields involving four $\P$'s, or three $\P$'s and one $\Pd$, in a footnote below.) {\it A priori}, there is a large number of possible terms to consider. However, for each integrand containing more than two derivatives acting on a single superfield, one can use integration by parts to get a sum of terms with at most two derivatives on any field. Similarly, for each term containing fewer than two derivatives acting on any superfield (in fact, the only such term is $D\Phi D\Phi \Db \Phi^\dag \Db \Phi^\dag$) we can use integration by parts to get a sum of terms, each containing a field on which two derivatives act. Hence, we need only consider terms which contain at least one field with two derivatives acting on it. All other terms, possibly after using integration by parts, can be obtained from linear combinations of these. This leaves eight terms:
\bea && D^2 \P \Db^2 \Pd \P \Pd \label{eight} \\ && D^2 \P \P \Db \Pd \Db \Pd \\ && \Db^2 \Pd \Pd D \P D \P \\ && \Db D \P D \Db \Pd \P \Pd \\ && \Db D \P \Db D \P \P^{\dag 2} \\ && D \Db \Pd D \Db \Pd \P^2 \\ && \Db D \P D \P \Db \Pd \Pd \\ && D \Db \Pd \Db \Pd D \P \P \ .
\eea
(Here and throughout this paper, derivatives are understood to act only on the nearest superfield. For example,  $D^2 \P \Db^2 \Pd \P \Pd = (D^2 \P) (\Db^2 \Pd) \P \Pd$.)
However, the last four terms can in fact be related to the first four using integration by parts, so really there are only four possible terms. Their superspace integrals yield the following Lagrangians when restricted to the complex scalar $A:$
\bea
D^2 \P \Db^2 \Pd \P \Pd \mid_{\th\th\tb\tb} &=&  16 A \As \Box A \Box \As \label{burt1}\\ D^2 \P \P \Db \Pd \Db \Pd \mid_{\th\th\tb\tb} &=&  16 A \Box A \pt \As \cdot \pt \As  \label{burt2} \\ \Db^2 \Pd \Pd D \P D \P \mid_{\th\th\tb\tb}  &=& 16 \As \Box \As \pt A \cdot \pt A \label{burt3}\\  \Db_{\ad} D^\a \P D_\a \Db^{\ad} \Pd \P \Pd \mid_{\th\th\tb\tb} &=& 8[A \As \Box A \Box \As + A \Box A \pt \As \cdot \pt \As + \As \Box \As \pt A \cdot \pt A \nn \\ && + (\pt A \cdot \pt A)(\pt \As \pt \As)] \label{burt4} \ .
\label{BuildingBlocks}
\eea
The real part of $(\pt A \cdot \pt A)(\pt \As \cdot \pt \As)$ reduces to $X^2$ and, hence, this is the term that we want \footnote{We also could have considered terms with three $\P$s and one $\Pd$, or terms with four $\P$s, plus their hermitian conjugates. Following the same logic as above, and making use of integration by parts, it is straightforward to see that the terms with four $\P$s (for example, $\Db D \P \Db D \P \P^2$) are all zero after integrating over $d^4\th$. For the case of three $\P$s and one $\Pd$, there are only two independent terms, $\Db^2 \Pd D^2 \P \P^2 \mid_{\th\th\tb\tb} = A^2 \Box A \Box \As$ and $\Db^2 \Pd D \P D \P \P \mid_{\th\th\tb\tb} = A (\pt A)^2 \Box \As,$ and since neither contains $X^2$ we do not consider these terms any further.}. We can isolate it by taking an obvious linear combination of the above terms; namely,
$- \frac{1}{16}\cdot [\eqref{burt1}+\eqref{burt2}+\eqref{burt3}]+\frac{1}{8} \cdot \eqref{burt4}$.
This linear combination can in fact be re-written, using integration by parts, as the single term $\frac{1}{16}D\P D\P \Db\Pd \Db \Pd.$ In terms of all of the component fields (up to quadratic order in the spinor $\psi$), it is given by
\bea
\frac{1}{16}D\P D\P \Db \Pd \Db \Pd &=& \th\th\tb\tb \Big[(\pt A)^2(\pt\As)^2-2\Fs F \pt A \cdot \pt \As + F^{\ast 2}F^2 \nn \\
&& \quad \quad \quad -\frac{i}{2} (\psi \s^\m\sb^\l \s^\n \bar\psi_{,\n})A_{,\m}A^\ast_{,\l}+\frac{i}{2}(\psi_{,\n}\s^\n\sb^\m \s^\l \bar\psi)A_{,\m}A^\ast_{,\l} \nn \\
&& \quad \quad \quad+ i \psi \s^\m \bar\psi^{,\n}A_{,\m}A^\ast_{,\n} -i \psi^{,\m}\s^\n \bar\psi A_{,\m} A^\ast_{,\n}+ \frac{i}{2} \psi \s^\m\bar\psi (A^\ast_{,\m} \Box A -A_{,\m}\Box \As)\nn \\
&& \quad \quad \quad+\frac{1}{2}(F\Box A-\pt F\pt A)\pb\pb + \frac{1}{2}(\Fs\Box \As - \pt \Fs \pt \As)\psi\psi  \nn \\
&& \quad \quad \quad+ \frac{1}{2}F A_{,\mu}(\pb \sb^\m \s^\n \pb_{,\n}-\pb_{,\n}\sb^\m \s^\n \pb)+\frac{1}{2}\Fs \As_{,\m}(\psi_{,\n}\s^\n \sb^\m \psi - \psi \s^\n \sb^\m \psi_{,\n})  \nn \\
&& \quad \quad \quad+\frac{3i}{2}\Fs F(\psi_{,\m}\s^\m \pb - \psi \s^\m \pb_{,\m}) + \frac{i}{2} \psi \s^\m \pb(F \Fs_{,\m} - \Fs F_{,\m})\Big] \nn \\
&& +\sqrt{2}i \tb\tb (\pt A)^2 \th\s^\m\bar\psi A^\ast_{,\m} -\sqrt{2}i \th\th (\pt \As)^2 \psi\s^\m \tb A_{,\m} \nn \\
&& +\sqrt{2} \th\th F \As_{,\m}(i\Fs \psi \s^\m \tb + \tb \sb^\n \s^\m \pb A_{,\n}) + \sqrt{2} \tb\tb \Fs A_{,\m}(-iF\th\s^\m \pb + \psi \s^\m \sb^\n \th \As_{,\n})\nn \\
&& -\frac{1}{2}\tb\tb (\pt A)^2 \bar\psi \bar\psi -\frac{1}{2}\th\th (\pt \As)^2 \psi\psi +2 (\psi\s^\m \tb)( \th \s^\n \bar\psi)A_{,\m} A^\ast_{,\n} \nn \\
&& + 2\Fs F \th\psi \tb \pb +i\th\s^\m\tb(F A_{,\m}\pb\pb - \Fs \As_{,\m}\psi\psi) + \frac{1}{2} \th\th F^2 \pb\pb + \frac{1}{2}\tb\tb F^{\ast 2}\psi\psi \nn \\
&& + \sqrt{2} \Fs F (\Fs \th\psi + F \tb\pb) + i \psi \s^\m \pb(F \As_{,\m}-\Fs A_{,\m}) \label{Xsquared} \label{new1} \\
&& \supset \th\th\tb\tb \frac{1}{4}(\pt\p)^4.
\eea
This superfield has a number of special properties:

1. It is a ``clean'' supersymmetric extension of $X^2,$ in the sense that this superfield contains $X^2$ and no other purely $\phi$-dependent terms. (Just to make this point clear: we did not want to obtain something like $\p\Box\p (\pt\p)^2,$ which, after integration by parts, would include $(\pt\p)^4,$ but would also generate additional unwanted terms.)

2. We might have expected to find a term containing $\left(\pt A \cdot \pt \As\right)^2,$ which, in terms of the real scalars $\p$ and $\chi,$ would have given the Lagrangian $\frac{1}{4}\left((\pt\p)^2 + (\pt\chi)^2\right)^2.$ However, this is not what supersymmetry provides. Instead one gets $(\pt A)^2 (\pt \As)^2,$ which corresponds to the Lagrangian
\be
(\pt A)^2 (\pt \As)^2 = \frac{1}{4}(\pt\p)^4 + \frac{1}{4}(\pt\chi)^4 - \frac{1}{2} (\pt\p)^2 (\pt\chi)^2 + (\pt\p\cdot\pt\chi)^2.
\label{new2}
\ee
An interesting feature is the last term, which mixes $\p$ and $\chi.$ Note that if we restrict our attention to purely time-dependent fields, the above Lagrangian reduces to the $O(2)$-invariant form $\frac{1}{4}\left(\dot\p^2 + \dot\chi^2\right)^2,$ as one would also obtain from
$(\pt A \cdot \pt \As)^2$. On the other hand, the dependence of the supersymmetric term \eqref{new2} on spatial gradients is more involved.

3. Although we will discuss the auxiliary field in more detail later on, here we simply point out that, despite it containing higher-derivatives, the superfield action \eqref{new1} does not generate a kinetic term for $F$.

4. With the spinor set to zero (which can be done consistently, as there are no terms linear in $\psi$), the only non-zero component of superfield \eqref{new1} is the top $\theta \theta {\bar{\theta}} {\bar{\theta}}$ component --- all lower components vanish! This property renders this term particularly useful as a building block for constructing larger terms that include $X^2,$ as we will elaborate on in the next section.

For completeness, we should add that there exists a second, inequivalent way of removing the unwanted terms from Eq. (\ref{burt4}), which involves subtracting off the term $\frac{1}{2}\P^2 \Db D \P D \Db \Pd$ plus its hermitian conjugate \cite{BaumannGreen}. The end result is proportional to $(\P-\Pd)^2 \Db D \P D \Db \Pd,$ and this term also represents a supersymmetric extension of $X^2.$ Up to quadratic order in fields other than $\p,$ its components read \bea -\frac{1}{16} (\P - \Pd)^2 \Db D \P D \Db \Pd &=&
\th\th\tb\tb [\frac{1}{4}(\pt\p)^4 + \frac{1}{4}(\pt\p)^2(\pt\chi)^2 - \frac{1}{4}(\pt\p)^2\chi \Box \chi - \phi_{,\m}\p_{,\n}\chi \chi^{,\m\n}
\nn \\ && \qquad + \phi_{,\m}\chi_{,\n}\chi \p^{,\m\n} + \frac{1}{4}\chi^2 \p^{,\m\n}\p_{,\m\n}-\frac{1}{4}\chi^2 \pt\phi \cdot \pt (\Box \p) \nn \\ &&
\qquad -\frac{i}{4}(\pt\p)^2(\psi_{,\m}\s^\m \pb - \psi \s^\m \pb_{,\m})-\frac{i}{2}\p_{,\m}\p_{,\n}(\psi^{,\m}\s^\n \pb - \psi \s^\n {\bar\psi}^{,\m})] \nn \\ &&
+i(\pt\p)^2\th\s^\m\tb \p_{,\m}(\th\psi-\tb\pb) + \frac{1}{2}(\pt\p)^2 \chi (\th\th \psi_{,\n}\s^\n \tb + \tb\tb \th \s^\n \bar\psi_{,\n})
\nn \\ && -2 \th\s^\m \tb \chi \p_{,\m}\p_{,\n}(\th\psi^{,\n}+\tb{\bar\psi}^{,\n}) -\frac{1}{\sqrt{2}}(\pt\p)^2(\th\th F \tb\pb + \tb\tb \Fs \th\psi) \nn \\ &&
-(\pt\p)^2\th\psi\tb\pb-(\pt\p)^2\th\s^\m\tb\chi \p_{,\m}
-\frac{1}{4}\th\th (\pt\p)^2\psi\psi -\frac{1}{4}\tb\tb(\pt\p)^2\pb\pb \nn \\ && + \frac{i}{\sqrt{2}}(\pt\p)^2(\th\th \chi F - \tb\tb \chi \Fs)
+i(\pt\p)^2\chi (\th\psi - \tb\pb)
-\frac{1}{2}(\pt\p)^2\chi^2. \label{XsquaredBG}
\eea
This superfield shares a number of properties with Eq. (\ref{Xsquared}): it is also a clean supersymmetric extension of $X^2,$ and does not contain a kinetic term for the auxiliary field $F.$ Moreover, up to the order considered above, the fermionic terms are identical for both supersymmetric extensions. This implies that, to a large extent, it does not matter which of the two superfield extensions of $X^2$ we use. However, the term (\ref{XsquaredBG}) does not reduce to its top component when the fermion is set to zero. This fact, together with the presence of numerous undifferentiated occurrences of the second scalar $\chi,$ renders this second version less useful in the construction of supersymmetric extensions of $X^n$ for $n>2,$ and we will thus use the former expression (\ref{Xsquared}) for the remainder of this work.

\subsection{$X^n$}
\label{Xn}

We have just seen that the supersymmetric extension of $X^2$ enjoys the property that this superfield, when the spinor is set to zero, contains only the top component proportional to $\th\th\tb\tb,$ and no lower components (this is possible, incidentally, because it is a vector superfield). Moreover, this top component contains the term $(\pt\p)^4.$ Hence, it is now easy to construct supersymmetric terms that contain $(\pt\p)^4$ as a factor, such as the $X^n$ terms. Using the supersymmetry algebra (\ref{susyalgebra}), we have
\be \frac{1}{32}\{D,\Db\}(\P + \Pd)\{D,\Db\}(\P+\Pd) = -\frac{1}{2}(\pt\p)^2 \, + \; {\text{higher components}}\,.
\ee
Hence, one can generate supersymmetric extensions of $X^n$, $n\geq2$ by considering the superfields
\be
X^n \subset \frac{1}{16}D \P D \P \Db \Pd \Db \Pd \left.\left[\frac{1}{32}\{D,\Db\}(\P + \Pd)\{D,\Db\}(\P + \Pd)\right]^{n-2}\right\vert_{\th\th\tb\tb}.
\ee
More precisely, up to quadratic order in the fields $\psi$ and $F$, we have
\bea
&& \frac{1}{16}D \P D \P \Db \Pd \Db \Pd \left.\left[\frac{1}{32}\{D,\Db\}(\P + \Pd)\{D,\Db\}(\P + \Pd)\right]^{n-2}\right\vert_{\th\th\tb\tb,\, {\rm quad.}}\nn \\
&&= \left[-\frac{1}{4}(\pt(A + \As))^2\right]^{n-2}\Big[(\pt A)^2(\pt \As)^2-2\Fs F \pt A \cdot \pt \As \nn \\
&& \qquad \qquad \qquad -\frac{i}{2} (\psi \s^\m\sb^\l \s^\n \bar\psi_{,\n})A_{,\m}A^\ast_{,\l}+\frac{i}{2}(\psi_{,\n}\s^\n\sb^\m \s^\l \bar\psi)A_{,\m}A^\ast_{,\l} \nn \\
&& \qquad \qquad \qquad+ i \psi \s^\m \bar\psi^{,\n}A_{,\m}A^\ast_{,\n} -i \psi^{,\m}\s^\n \bar\psi A_{,\m} A^\ast_{,\n}+ \frac{i}{2} \psi \s^\m\bar\psi (A^\ast_{,\m} \Box A -A_{,\m}\Box \As)\Big] \nn \\
&& + \frac{n-2}{2}\left[-\frac{1}{4}(\pt(A + \As))^2\right]^{n-3}(A+\As)_{,\m}(\As_{,\l}(\pt A)^2 i \psi^{,\m}\s^\l \pb - A_{,\l}(\pt \As)^2 i \psi \s^\l \pb^{,\m}) \nn \\
&& + \frac{n-2}{4}\left[-\frac{1}{4}(\pt(A + \As))^2\right]^{n-3} A_{,\l}\As_{,\n}(A^{,\t}+A^{\ast,\t})(A_{,\t\m}-\As_{,\t\m})i\psi\s^\l\sb^\m\s^\n\pb\,.
\eea
Writing $A$ in terms of the real scalars $\p$ and $\chi$, and restricting to quadratic order in $\chi$, this becomes
\bea
&& \frac{1}{16}D \P D \P \Db \Pd \Db \Pd \left.\left[\frac{1}{32}\{D,\Db\}(\P + \Pd)\{D,\Db\}(\P + \Pd)\right]^{n-2} \right\vert_{\th\th\tb\tb,\, {\rm quad.}} \nn \\
&& = X^n +X^{n-1}(\pt\chi)^2 + X^{n-2}(\pt\p\cdot\pt\chi)^2 + 2 X^{n-1}\Fs F\nn \\
&& +X^{n-1} \frac{i}{2}(\psi_{,\m}\s^\m \bar\psi - \psi \s^\m \bar\psi_{,\m}) - (n-1)X^{n-2}\p_{,\m}\p_{,\n}\frac{i}{2} (\psi^{,\n}\s^\m \bar\psi - \psi \s^\m \bar\psi^{,\n})\,.
\label{SusyRho}
\eea

There are closely related terms which are also of interest, namely
\bea
&& \frac{1}{16^2}D \P D \P \Db \Pd \Db \Pd \left[\{D,\Db\}(\P - \Pd)\{D,\Db\}(\Pd - \P)\right] \Big\vert_{\th\th\tb\tb, \, {\rm quad.}} \nn \\
&& = -X^2(\pt\chi)^{2}, \label{SusyChi}\\ && \frac{1}{16^3}D \P D \P \Db \Pd \Db \Pd \left[\{D,\Db\}(\P + \Pd)\{D,\Db\}(\P - \Pd)][\{D,\Db\}(\P + \Pd)\{D,\Db\}(\Pd - \P)\right] \Big\vert_{\th\th\tb\tb, \, {\rm quad.}} \nn \\
&& = X^2(\pt\p\cdot\pt\chi)^2\,.
\label{SusyMixed}
\eea
These terms do not contribute to $P(X,\phi)$. However, they have the feature that, from a model-building point of view, they can change the properties of fluctuations of the $\chi$ field independently from of those of $\p$ (and $\psi$) when $\p$ develops a time- or space-dependent vev. We will return to this point when discussing the ghost condensate background in Sec.~\ref{sectionGhostCondensate}.

Note also that one can take combinations of (\ref{SusyRho}) and (\ref{SusyChi}) which resum to the terms
\bea && \frac{1}{16}  D\P D\P \Db \Pd \Db \Pd \left(\frac{1}{8}\Db D \P D \Db \Pd\right)^{n-2}\Big\vert_{\th\th\tb\tb} \supset (\pt A)^2(\pt \As)^2(-\pt A \cdot \pt \As)^{n-2}\,.
\label{SusySymmetric}
\eea
Such terms allow us to obtain a supersymmetric extension that is symmetric in terms of $\p$ and $\chi,$ as it consists of terms of the form
\be
\left[\frac{1}{4}(\pt \phi)^4 + \frac{1}{4}(\pt \chi)^4 -\frac{1}{2}(\pt \phi)^2(\pt \chi)^2 + (\pt\phi \cdot \pt \chi)^2\right]\cdot \left[-\frac{1}{2}(\pt \phi)^2 -\frac{1}{2} (\pt \chi)^2\right]^{n-2}\,.
\ee
Moreover, in terms of time dependence alone, this gives the simple extension
\be
P\left(\frac{1}{2}\dot\p^2\right) \rightarrow P\left(\frac{1}{2}\dot\p^2 + \frac{1}{2} \dot\chi^2\right)\,,
\ee
which preserves the functional form of $P$.

\subsection{The Supersymmetric Extension of $P(X,\phi)$}
\label{Pfull}

So far, we have only shown how to construct supersymmetric versions of $X^n.$ It is straightforward, however, to multiply these terms by field-dependent coefficients. Note that
\be
\left[\frac{1}{\sqrt{2}}(\P + \Pd)\right]^k = \p^k + \sqrt{2} k \p^{k-1} (\th\psi + \tb \pb) - k \p^{k-1}\th\s^\m\tb \chi_{,\m} + \cdots
\ee
Now consider
\bea
&& \left[\frac{1}{\sqrt{2}}(\P + \Pd)\right]^k \frac{1}{16}D\P D \P \Db \Pd \Db \Pd \Big\vert_{\th\th\tb\tb} \nn \\ &&= \p^k (X^2 + \cdots)  \nn \\ && -\frac{k}{4} \p^{k-1} \psi \s^\n \sb^\m \s^\l \pb \p_{,\n} \p_{,\l} \chi_{,\m} -\frac{k}{\sqrt{2}}\p^{k-1}\psi \s^\m \pb (\pt\p)^2 \chi_{,\m} + \cdots
\eea
Hence, the only effect of multiplying the superfield \eqref{new1} by $[(\P + \Pd)/\sqrt{2}]^k$ is to multiply all terms that either do not contain $\chi$, $\psi$ and $F$, or are at most quadratic in these fields, by $\p^k$ . All other contributions are cubic and higher-order in fields other than $\p$ (we only wrote out the two leading terms). We do not need the higher-order terms for our purposes. However, it is straightforward to work them out should they be required. The same argument will go through for the terms (\ref{SusyRho}) when they are multiplied by $[(\P + \Pd)/\sqrt{2}]^k$. Once again, to quadratic order in fields other than $\p,$ they simply get multiplied by $\p^k.$

Hence, it is now clear that a supersymmetric extension of
\be
S_P = \int {\rm d}^4 x\, P(X,\p) = \int {\rm d}^4 x \sum_{n\in\mathbb{N}^\ast} a_n(\p)X^n
\ee
is given by
\bea
S_P^{\rm SUSY}&=& \int {\rm d}^4 x {\rm d}^4 \th K(\P,\Pd) \nn \\
&& \hspace{-1.5cm} +  \int {\rm d}^4 x {\rm d}^4 \th \sum_{n \geq 2} a_n\left(\frac{\P + \Pd}{\sqrt{2}}\right)\frac{1}{16}D \P D \P \Db \Pd \Db \Pd \left[\frac{1}{32}\{D,\Db\}(\P + \Pd)\{D,\Db\}(\P + \Pd)\right]^{n-2}\,,
\label{SusyPofX}
\eea
where $a_n\left(\frac{\P + \Pd}{\sqrt{2}}\right)$ is the same function as $a_n(\p).$ For the standard kinetic term $X$, we have written the supersymmetric generalization as the integral of the K\"ahler potential $K(\P,\Pd)$. This leads to the action $-a_{1}(\phi) \pt A \pt \As$, where $a_1(\phi)=K_{,A\As}\mid_{\th=\tb=0}$.
To quadratic order in fields other than $\p,$ action (\ref{SusyPofX}) reduces to
\bea
S_{P,\;{\rm quad.}}^{\rm SUSY} =&& \int {\rm d}^4 x \, a_1(\phi) \left[X - \frac{1}{2}(\pt\chi)^2 + \Fs F + \frac{i}{2}(\psi_{,\m}\s^\m \bar\psi - \psi \s^\m \bar\psi_{,\m})\right]\nn \\
&& \quad +\sum_{n \geq 2} a_n(\p)\Bigg[X^n +X^{n-1}(\pt\chi)^2 + X^{n-2}(\pt\p\cdot\pt\chi)^2 + 2X^{n-1} \Fs F \nn \\
&& \quad \qquad +X^{n-1} \frac{i}{2}(\psi_{,\m}\s^\m \bar\psi - \psi \s^\m \bar\psi_{,\m}) - (n-1)X^{n-2}\p_{,\m}\p_{,\n}\frac{i}{2} (\psi^{,\n}\s^\m \bar\psi - \psi \s^\m \bar\psi^{,\n})\Bigg]\,.
\label{SusyComponents}
\eea
It is important to realize that this action is not the unique supersymmetric extension of $P(X,\phi).$ For example, certain terms, such as (\ref{SusyChi}) and (\ref{SusyMixed}), can be added since they change the resulting dependence on the second real scalar $\chi,$ but do not change the purely $\p$-dependent terms.

Following up on our discussion at the end of the last section, we can make use of the terms (\ref{SusySymmetric}) to write a supersymmetric extension for which the derivative terms are symmetric in $\p$ and $\chi$. This is given by
 \bea
 S_P^{{\rm SUSY},\;{\rm symmetric}}&=& \int {\rm d}^4 x {\rm d}^4 \th K(\P,\Pd) \nn \\
 &&+ \int {\rm d}^4 x {\rm d}^4 \th \sum_{n \geq 2} a_n\left(\frac{\P + \Pd}{\sqrt{2}}\right)\frac{1}{16}D \P D \P \Db \Pd \Db \Pd \left(\frac{1}{8}\Db D \P D \Db \Pd\right)^{n-2}. \label{SusyPofXSymmetric}
 \eea
 This action has the same dependence (up to quadratic order) on $F$ and $\psi$ as (\ref{SusyPofX}) above.

Let us briefly discuss the auxiliary field $F$. In (\ref{SusyPofX}) there are some interaction terms that are quadratic in the spinor $\psi$ and involve derivatives on $F$ which cannot be removed using integration by parts (an example is provided by the final top-component term coming from (\ref{Xsquared})). However, there is no pure kinetic term for $F$, nor any term such as $(\pt A \cdot \pt \As)(\pt F \cdot \pt \Fs),$ which would act as a kinetic term for $F$ in a background with a non-zero vev of $(\pt\p)^2.$ This is fortunate, since it implies that $F$ has not become a propagating field. In order to show more explicitly how to eliminate the auxiliary field, let us consider the supersymmetric version of $X+b X^2$ for some dimensionless constant $b$ (a similar analysis will apply to the $X^n$ terms with $n>2$). Then the equation of motion for $F$ is
\bea
&&F = \frac{b}{m^4}\Bigg[2 F \pt A \cdot \pt \As - 2 \Fs F^2 - \Box \As \psi\psi - \frac{1}{2} \pt \As \cdot \pt(\psi\psi) - \frac{1}{2} \As_{,\m}(\psi_{,\n}\s^\n\sb^\m\psi - \psi \s^\n \sb^\m \psi_{,\n}) \nn \\
&&\qquad~~~~~~- \frac{3i}{2}F(\psi_{,\m}\s^\m \pb - \psi \s^\m \pb_{,\m})+F_{,\m}i\psi\s^\m\pb + \frac{i}{2}(\psi \s^\m \pb)_{,\m}F\Bigg]\,.
\eea
We can solve this equation perturbatively as a series in $b/m^4$ by writing $F \equiv F_0 + F_1 + \cdots $ where $F_i$ is at order $(b/m^4)^{i}$. To lowest order, $F_0=0$. At the next order, we find
\be
F_1 = \frac{b}{m^4}\left[- \Box \As \psi\psi - \frac{1}{2} \pt \As \cdot \pt(\psi\psi) - \frac{1}{2} \As_{,\m}(\psi_{,\n}\s^\n\sb^\m\psi - \psi \s^\n \sb^\m \psi_{,\n})\right]\,.
\ee
When plugged back into the action, this leads to the terms \be \Delta {\cal L} = \frac{b^2}{m^8}\Box A \Box \As \psi \psi \pb \pb + \cdots, \ee which are all corrections to the higher-order interaction terms. These do not alter the vacuum or stability properties of the theory and, hence, we need not consider them for our analysis.

Even though this goes beyond the scope of the present paper, we note that in the presence of a superpotential $W,$ the elimination of the auxiliary field leads to interesting new terms. In that case, the lowest order solution is $F_0 = - \pt W^\ast/\pt \As$, as usual. At the next order, we find
\bea
F_1 &=& \frac{b}{m^4}\Bigg[- \Box \As \psi\psi - \frac{1}{2} \pt \As \cdot \pt(\psi\psi) - \frac{1}{2} \As_{,\m}(\psi_{,\n}\s^\n\sb^\m\psi - \psi \s^\n \sb^\m \psi_{,\n})  -2 \frac{\pt W^\ast}{\pt \As} \pt A \cdot \pt \As \nn \\ &&+ 2 \frac{\pt W}{\pt A}(\frac{\pt W^\ast}{\pt \As})^2  + \frac{3i}{2}\frac{\pt W^\ast}{\pt \As}(\psi_{,\m}\s^\m \pb - \psi \s^\m \pb_{,\m})-\frac{\pt W^\ast}{\pt \As}_{,\m}i\psi\s^\m\pb - \frac{i}{2}(\psi \s^\m \pb)_{,\m}\frac{\pt W^\ast}{\pt \As}\Bigg]\,.
\eea
Plugging this back into the action leads to corrections at order ${\cal O}(b/m^4)$ of the form (for example)
 \be \frac{b}{m^4}V \pt A \cdot \pt \As\,,\quad \frac{b}{m^4}Vi(\psi_{,\m}\s^\m \pb - \psi \s^\m \pb_{,\m})\,,\quad \frac{b}{m^4} V^2\,,
 \ee
 where the lowest order potential is $V(A,\As)= \frac{\pt W}{\pt A}\frac{\pt W^\ast}{\pt \As}.$ In other words, there are corrections to both the kinetic and potential terms, proportional to the potential itself. This may lead to interesting new effects for models where the potential contributes significantly to the dynamics.

As a final comment: above, we solved the equation for $F$ as a series expansion around the usual solutions; that is, we assumed that the higher-derivative kinetic terms provide a suitably small correction. However, the equation for $F$ is cubic, and there may be other branches of solutions that are not continuously connected to the usual solutions when the higher-derivative terms are switched off. This may lead to an entirely new role for the ``auxiliary'' field $F$. We leave an investigation of this subject to future work.

\section{An Example: Supersymmetric DBI Actions}
\label{egDBI}

We can immediately apply our formalism to write supersymmetric versions of DBI actions of the form
\be {\cal L}_{\rm DBI} = -\frac{1}{f(\p)}\left(\sqrt{1+f(\p)(\pt\p)^2}-1\right)\,.\label{DBIaction}
\ee
The prototypical DBI theory describes a D3-brane moving radially (with $\p$ parameterizing the radial position of the brane) in $AdS_5$ space in the context of the $AdS_5 \times S^5$ compactification of type IIB string theory; in that case~\cite{Silverstein:2003hf} $f(\p) \propto \p^{-4},$ but we will leave the function $f$ arbitrary here. In fact, the string theoretic origin of the action (\ref{DBIaction}) is irrelevant to this paper -- here, our goal is simply to show how to obtain linearly realized ${\cal N}=1$ supersymmetric extensions of it. (Which particular such supersymmetric extension will arise in a given string theoretic context will depend on all the details of the specific compactification considered.) In terms of $X$, we can re-write the action as the series
\be  {\cal L}_{\rm DBI} = X - \sum_{n\geq 2} {\frac{1}{2} \choose n} f^{n-1}(-2X)^n\,,
\ee
and hence we can immediately obtain a supersymmetric version of this action by inserting into~(\ref{SusyPofX}) the relations $K=\Pd \P$ and
\be
a_{n\geq 2}=-f\left(\frac{\P + \Pd}{\sqrt{2}}\right)^{n-1}{\frac{1}{2} \choose n} (-2)^n\,.
\ee
In components, up to quadratic order in fields other than $\p$, this yields~(\ref{SusyComponents}) with $a_1=1$ and $a_{n\geq 2}$ as just given.

A different supersymmetric version of the DBI action, which treats the $\phi$ and $\chi$ kinetic terms more symmetrically, can be obtained using (\ref{SusyPofXSymmetric}), giving
\bea {\cal L}_{\rm DBI}^{\rm SUSY} &=& \int {\rm d}^4\th \Pd \P - \frac{1}{4}\sum_{n \geq 2} {\frac{1}{2} \choose n}
f\left(\frac{\P + \Pd}{\sqrt{2}}\right)^{n-1}D \P D \P \Db \Pd \Db \Pd \left(-\frac{1}{4}\Db D \P D \Db \Pd\right)^{n-2}\,.
\eea
Looking at time-dependence only, with the auxiliary field set to zero and where we write out only the interactions of the fermion with $\p$ (up to quadratic order in fermions), this re-sums to the Lagrangian
\bea {\cal L}_{\rm DBI}^{\rm SUSY} &=& -\frac{1}{f(\p)}\left(\sqrt{1-f(\p)(\dot\p^2+\dot\chi^2)}-1\right) \nn \\ && +i(\psi_{,0}\s^0 \pb - \psi \s^0 \pb_{,0})\left[\sum_{n \geq 1}(2n-1){\frac{1}{2} \choose n}f(\p)^{n-1}(-\dot\p^2)^{n-1}\right] \nn \\ &=& -\frac{1}{f(\p)}\left(\sqrt{1-f(\p)(\dot\p^2+\dot\chi^2)}-1\right) \nn \\ && +i(\psi_{,0}\s^0 \pb - \psi \s^0 \pb_{,0}) \frac{1}{f(\p)\dot\p^2}\left[ \frac{1}{\sqrt{1-f(\p)\dot\p^2}} - 1 \right]\,,
\eea
where we have used
\be
\left(\frac{1}{2}-n\right){\frac{1}{2} \choose n} =\left(\frac{1}{2}-n\right)\frac{\frac{1}{2}!}{n!\left(\frac{1}{2}-n\right)!}=\frac{\frac{1}{2}\left(-\frac{1}{2}\right)!}{n!(-\frac{1}{2}-n)!}= \frac{1}{2} {-\frac{1}{2} \choose n}\,.
\ee
Without pursuing this topic in the present paper, we simply note that, interestingly, the characteristic relativistic factor $\sqrt{1-f\dot\p^2}$ has moved to the denominator in the fermionic term.

\section{The Supersymmetric Ghost Condensate} \label{sectionGhostCondensate}

Arguably, the most surprising property of $P(X)$ theories is the fact uncovered in \cite{ArkaniHamed:2003uy} that, in these theories, the NEC can be violated in a controlled way under certain circumstances. This is an astonishing result, since NEC violations are usually considered to be pathological. They tend to be associated with the appearance of ghost fields, that is,  physical fluctuations with the wrong-sign kinetic term, which render the vacuum unstable.
The way in which this fate is circumvented for $P(X)$ theories is through a combination of higher derivative effects and because such theories admit a time-dependent, Lorentz symmetry-breaking vacuum --- the {\it ghost condensate} --- which we review below.

We are interested in studying the supersymmetric version of the ghost condensate theory, for a number of reasons. First of all, it should be interesting just to see what we get! Higher-derivative scalar theories have, to our knowledge, not been studied in the context of minimal 4-dimensional supersymmetry. One aspect of this work is simply to show how such supersymmetric models can be constructed. Secondly, it will be interesting to see what effects the NEC violation has on the various additional fields, that is, $\chi$, $\psi_{\alpha}$ and $F$, that are required by supersymmetry. One might expect anything from no effect at all, to the extra fields spoiling the viability of the model altogether. Thirdly, it is not clear at present whether or not the ghost condensate model can be obtained as an effective theory from a string theoretic setup~\cite{Dubovsky:2006vk}. Reformulating the model in a supersymmetric fashion may help in elucidating this question. And finally, NEC violations are required to describe certain types of dynamics in a cosmological context at high energy. Examples include non-singular cosmological bounces (such as those considered in New Ekpyrotic models \cite{Buchbinder:2007ad,Buchbinder:2007tw}), as well as the dynamics of the internal dimensions in Kaluza-Klein models of inflation (see \cite{Wesley:2008de,Wesley:2008fg,Steinhardt:2008nk}). Since supersymmetry may  be a relevant symmetry of nature at high energies, this provides further motivation for supersymmetrizing the effective ghost condensate theory.

\subsection{A Brief Review of Ghost Condensation}

For a Lagrangian ${\cal L} = P(X),$ the equation of motion is given by
\be \frac{{\rm d}}{{\rm d}t}\left(P_{,X}\dot\phi\right)=0\,,
\ee
where we assume that the solution does not depend on spatial position. Typically, one takes $\p={\rm constant},$ which is evidently a solution of the above equation. However, the equation of motion also allows for solutions with constant $X$. These corresponds to having a vev of $\p$ that grows linearly with time,
\be \p = ct,
\label{wind1}
 \ee
 where $c$ is a constant. Any constant will do. However, for future reference, we bear in mind that in a cosmological context, the equation of motion becomes
\be \frac{{\rm d}}{{\rm d}t}\left(a^3 P_{,X}\dot\phi\right)=0,
\ee
where $a(t)$ is the scale factor of the universe. This equation implies that Hubble friction drives the evolution close to a local extremum of $P(X)$ where $P_{,X}=0.$

The explicit time dependence of solution \eqref{wind1} breaks Lorentz invariance, and this leads to interesting properties. First of all, evaluating the energy and pressure densities one finds
\bea
\rho &=& 2XP_{,X} -P\,; \nn \\ p &=& P\,,
\eea
so that $\rho + p = 2X P_{,X}.$ Since by definition $X>0$, the NEC can be violated if $P_{,X}<0$. That is, if we are close to an extemum of $P(X)$, then on one side the NEC is satisfied while on the other it is not. For small fluctuations $\d\p(t,\vec{x})$ around the solution \eqref{wind1}, the Lagrangian can be expanded to quadratic order in derivatives as
\be {\cal L}_{\rm quad.} = \frac{1}{2}\left[ (2XP_{,XX}+ P_{,X})(\d\dot\p)^2 - P_{,X}\d\p^{,i}\d\p_{,i}\right]\,.
\label{expansionaroundextremum}
\ee
As a result of Lorentz-breaking, the coefficients in front of the time and space pieces are unequal. We see that the condition for the absence of a ghost is
\be 2X P_{,XX} + P_{,X} >0\,,
\label{gcstability}
\ee
which is automatically satisfied close to a local {\it minimum} of $P(X)$ and, importantly, for some distance into the NEC violating region itself.  Hence, in this theory, the {\it NEC can be violated without the appearance of a ghost field}. The price one pays is that the spatial gradient term, proportional to $P_{,X},$ is now very small and, more worryingly, has the wrong sign when the NEC is violated. This signals a gradient instability whose effects, happily, can be mitigated by introducing additional higher-derivative terms, such as $-(\Box \phi)^2$,  into the Lagrangian.  Due to the smallness of $P_{,X}$ in the regime of interest, such terms do indeed become relevant, and they imply that the model can remain stable over a certain timescale. In cosmology, if one wants to consider theories in which the universe transitions, {\it  in a smooth way}, from a pre-big bang contracting phase to the standard expanding phase, then during this bounce {\it the NEC must be violated}. If the cause of NEC violation is a ghost condensate, then a constraint on the model is that the timescale over which the gradient instability is harmless must be at least as long as it takes for the universe to bounce~\cite{Creminelli:2006xe}. See, for example, \cite{Creminelli:2006xe,Buchbinder:2007ad} for a detailed stability analysis.

\subsection{The Supersymmetric Ghost Condensate}

The ghost condensate arises as a solution of the equation of motion close to a local minimum of $P(X)$, for a positive value of $X$. By expanding around the minimum and re-scaling the field $\phi$ so that the minimum lies at $X=\frac{1}{2}$ (this corresponds to choosing $c=1$), one can write the prototypical ghost condensate action as
\be
\int {\rm d}^4 x \left[-X+X^2\right] = \int {\rm d}^4x \left[+\frac{1}{2}(\pt \phi)^2 + \frac{1}{4}(\pt \phi)^4\right] \,.
\label{ActionGC}
\ee

We now want to supersymmetrize this action. Using \eqref{rain2} and \eqref{Xsquared}, we obtain (setting $F=0$ and up to quadratic order in $\chi$ and $\psi$)
 \bea
  \nn
 {\cal L}^{\rm SUSY}&=& \left[-\Pd\P +\frac{1}{16}D\P D\P \Db \Pd \Db \Pd\right] \Big\vert_{\th\th\tb\tb}  \\
  \nn
 &=& +\frac{1}{2}(\pt\p)^2 + \frac{1}{4}(\pt\p)^4 +\frac{1}{2} (\pt\chi)^2 - \frac{1}{2}(\pt\p)^2(\pt\chi)^2 + (\pt\p\cdot\pt\chi)^2  \label{gcSusic} \\
 & - &\frac{i}{2}(\psi_{,\m}\s^\m \bar\psi - \psi \s^\m \bar\psi_{,\m}) -\frac{1}{2}(\pt\p)^2\frac{i}{2}(\psi_{,\m}\s^\m \bar\psi - \psi \s^\m \bar\psi_{,\m}) -\p_{\m}\p_{,\n}\frac{i}{2}(\psi^{,\n}\s^\m \bar\psi - \psi \s^\m \bar\psi^{,\n}).
 \eea
The bosonic equations of motion are
\bea \Box \phi  &=&  \Box \phi (\chi_{,\mu}\chi_{,\nu}-\phi^{,\mu}\phi_{,\mu})-2\phi^{,\mu\nu}(\phi_{,\mu}\phi_{,\nu}+\chi_{,\mu}\chi_{,\nu})
-2\Box\chi\chi^{,\mu}\phi_{,\mu} \,, \\ \Box \chi &=& \Box \chi (\phi_{,\mu}\phi_{,\nu}-\chi^{,\mu}\chi_{,\mu})-2\chi^{,\mu\nu}(\chi_{,\mu}\chi_{,\nu}+\phi_{,\mu}\phi_{,\nu})
-2\Box\phi\phi^{,\mu}\chi_{,\mu} \,.
\eea
These can be solved with either one or two ghost condensates; that is, the general solution is of the form
\be
\phi = ct ~~, ~~ \chi = dt\,.
 \ee
However, {\it without loss of generality}, one can set $d=0$. To see this, note that under the normalized field redefinitions $\phi'=(c\phi + d\chi)/(c^2 + d^2)^{1/2}$ and $\chi'=(d\phi-c\chi)/(c^2 + d^2)^{1/2}$,
the $\phi'$ field corresponds purely to the ghost condensate while $\chi'$ vanishes. The fermion classical solution, of course, is zero.
To assess the stability of the supersymmetric ghost condensate, we can expand in small fluctuations around the background as (now setting $c=1$)
\be \phi = t + \d\phi(t,\vec{x})\,, \qquad \chi = \d\chi(t,\vec{x})\,, \qquad \psi=\d\psi(t,\vec{x}) \,.
\label{book1}
\ee
The result, to quadratic order, is
\bea
{\cal L}^{\rm SUSY}_{\rm quad.} &=& (\dot{\d\p})^2 + \, 0 \cdot \d\p^{,i}\d\p_{,i} \nn \\ && + \, 0 \cdot (\dot{\d\chi})^2 + \d\chi^{,i}\d\chi_{,i} \nn \\ && + \frac{1}{2} \frac{i}{2}\left(\d\psi_{,0}\s^0 \d\bar\psi - \d\psi \s^0 \d\bar\psi_{,0}\right)  -\frac{1}{2}\frac{i}{2}\left(\d\psi_{,i}\s^i \d\bar\psi - \d\psi \s^i \d\bar\psi_{,i}\right)\,.
\label{fluctuations}
\eea
The first line reproduces the standard result \eqref{expansionaroundextremum} for the single $\phi$ field ghost condensate, as it must. That is --- the time derivative term is ghost-free but, at the minimum of $P(X),$ the spatial gradient term for $\d\p$ vanishes. Furthermore, it would have a small coefficient with the {\it wrong sign} if we went into the NEC violating region, as discussed below Eq. (\ref{gcstability}). This gradient instability can be cured by including other higher derivative terms in the Lagrangian (that go outside of the $P(X)$ framework), such as $-(\Box\p)^2.$ The most obvious superfield expression containing this term in its highest component is
\bea
-\frac{1}{16}D^2 \P \Db^2 \Pd \Big\vert_{\th\th\tb\tb} &=& - \Box A \Box \As +  \pt \Fs \cdot \pt F - \frac{i}{2}\psi_{,\m}\s^\m\Box\bar\psi + \frac{i}{2} \Box \psi \s^\m \bar\psi_{,\m}\,.
\label{gradientstabilizer1}
\eea
The trouble with this expression is that it also includes a kinetic term for $F$, thus rendering the auxiliary field dynamical! However, other supersymmetric terms, which do not have this drawback, can fulfill the role of stabilizing the $\phi$ gradient. The simplest such example is
\be
-\frac{1}{2^{11}}D\P D \P \Db \Pd \Db \Pd \left[\{D,\Db\}\{D,\Db\}(\P + \Pd)\right]^2 \Big\vert_{\th\th\tb\tb,\;{\rm quad.}} =-(\Box \d\p)^2\,,
\label{gradientstabilizer2}
\ee
where we have evaluated this up to quadratic order in fluctuations in a ghost condensate background. To that order, \eqref{gradientstabilizer2} does not contain the auxiliary field $F$ (or $\chi$ and $\psi$) at all.

Having seen how the gradient instability of $\p$ can be addressed in a manner similar to the usual non-supersymmetric case, we now turn our attention to the new fields introduced by supersymmetry. As is evident from~(\ref{fluctuations}), the fluctuations of the second scalar $\chi$ come out entirely wrong. First, there is no time-derivative term for $\chi$ (and in the NEC-violating region, the coefficient becomes ghost-like) while the spatial gradient term is of the wrong sign. Taken at face value, this would be a serious instability of the model. It would essentially ruin the viability of the supersymmetric ghost condensate --- at least in the most interesting, NEC-violating regime. We show in the  Appendix that this bad behavior of $\chi$ is due entirely to the fact that $\p$ and $\chi$ enter (\ref{gcSusic}) in a symmetric way, and that this problem cannot be cured by looking at a deformed vacuum for one or both of the scalars.  It follows that  additional higher-derivative terms that are symmetric in $\p$ and $\chi$, such as (\ref{SusySymmetric}), will not resolve this problem. However, as soon as we break the symmetry between $\p$ and $\chi$, the $\chi$ fluctuations can be stabilized. With this in mind, consider the supersymmetric terms in (\ref{SusyChi}) and (\ref{SusyMixed}), each of which violates this symmetry. For example, the linear combination $8\cdot$(\ref{SusyChi})$-4\cdot$(\ref{SusyMixed}) yields
\bea
&& \frac{8}{16^2}D \P D \P \Db \Pd \Db \Pd \left[\{D,\Db\}(\P - \Pd)\{D,\Db\}(\Pd - \P)\right] \Big\vert_{\th\th\tb\tb, \, {\rm quad.}} \nn
\\ &&- \frac{4}{16^3}D \P D \P \Db \Pd \Db \Pd \left[\{D,\Db\}(\P + \Pd)\{D,\Db\}(\P - \Pd)\right]\left[\{D,\Db\}(\P + \Pd)\{D,\Db\}(\Pd - \P)\right] \Big\vert_{\th\th\tb\tb, \, {\rm quad.}} \nn \\
&&=-2(\pt\p)^4(\pt\chi)^2 - (\pt\p)^4(\pt\p\cdot\pt\chi)^2 \,.
 \label{walk1}
\eea
Adding this to  Lagrangian (\ref{gcSusic}), and expanding to quadratic order around the ghost condensate, changes both the time and spatial gradients of $\chi$ in \eqref{fluctuations} to
\be
{\cal L}^{\rm SUSY}_{\rm quad.}=\dots  +({\d\dot\chi})^2 -\d\chi^{,i}\d\chi_{,i} +\dots
\label{tonight1}
\ee
This renders the $\chi$ fluctuations stable, without changing anything else. Whether such terms would arise in a more fundamental derivation of the ghost condensate superfield action remains, of course, to be seen. However, from a purely effective field theory standpoint there is
{\it no instability} associated with the second scalar field $\chi$.

Finally, let us examine the fluctuations of the spinor $\psi$ around the ghost condensate. We see from \eqref{fluctuations} that, although the magnitudes of the two coefficients are equal, the time-derivative term is ghost-free while the spatial gradient term has the wrong sign \footnote{Had we used the superfield extension (\ref{XsquaredBG}) instead of (\ref{Xsquared}), we would have found exactly the same fermionic fluctuations, and hence also the wrong sign spatial gradient term.}.
Note that this is not the same kind of gradient instability as occurs for $\p$. There, the coefficient of the spatial derivative term is zero or small and, hence, higher-derivative terms can play a role in guaranteeing stability over an extended time period. For $\psi$, on the other hand, the coefficient of the wrong-sign spatial gradient term is not small. It follows that the inclusion of higher-derivative terms, such as those in (\ref{gradientstabilizer1}), is necessarily irrelevant. The situation for the fermion, therefore, is more akin to that of the second scalar $\chi$, whose deep wrong-sign spatial gradient had to be corrected by the addition of a new second order term --- the sum of the two kinetic spatial gradients having the correct sign. However, within the context of the superfield expressions we have analyzed --- with the same number of fields and derivatives as the $X^n$ terms ---
we are unable to find a fermionic analog of this mechanism. That is, within the context of the supersymmetric extension of the {\it pure $P(X)$ theory}, the fermion kinetic {\it spatial gradient} term has the wrong sign!

Be this as it may, it is unclear to us whether this is actually an instability. First of all, unlike the scalar fields where a wrong sign kinetic term destabilizes the {\it classical} vacuum, the background classical fermion is necessarily zero. Furthermore, as discussed in the next subsection, the supersymmetric ghost condensate spontaneously breaks supersymmetry and, hence, $\psi$ is a Goldstone fermion. When coupled to supergravity, one expects $\psi$ to be ``eaten'' by the super-Higgs mechanism and removed from the massless spectrum. Finally, an instability could arise by spontaneous fermion creation from the vacuum. However, showing this actually occurs would require a consistent quantization of such wrong-gradient fermions on a curved background and a study of their back-reaction on the geometry. For these reasons, we are, at the moment, agnostic about whether or not the wrong-sign spatial fermion kinetic term is a physical problem. A study of all of the above issues is underway.

We want to emphasize that if one is prepared to expand the discussion to supersymmetric extensions of Galileon scalar field theories, then, in addition to retaining the good kinetic behavior of the scalar fields, both the temporal and spatial parts of the fermion kinetic energy can easily be made to have the correct sign. That is, the fermion is both ghost-free and without a gradient instability! A proof of this result, and a complete exposition of supersymmetric Galileon theory, will be presented in a forthcoming publication.

For completeness, we present the entire supersymmetric extension of the $P(X)$ ghost condensate theory, combining all of the terms discussed independently above. The result is
\bea
{\cal L}_{\rm Ghost\; Condensate}^{\rm SUSY} &=& -\Pd \P \mid_{\th\th\tb\tb} \,  + \frac{1}{16}D\P D \P \Db \Pd \Db \Pd \mid_{\th\th\tb\tb} \nn \\
&& + D\P D \P \Db \Pd \Db \Pd \Bigg[ -\frac{1}{2^{11}}\left[\{D,\Db\}\{D,\Db\}(\P +\Pd)\right]^2 \nn \\ && \qquad \qquad \qquad \qquad + \frac{1}{2^5}\{D,\Db\}(\P-\Pd)\{D,\Db\}(\Pd-\P) \nn \\ && \qquad \qquad \qquad \qquad - \frac{1}{2^{10}}\left[\{D,\Db\}(\P+\Pd)\{D,\Db\}(\P-\Pd)\right]^2\Bigg] \Bigg\vert_{\th\th\tb\tb}\,,
\eea
In components, writing out all the terms that are relevant for a stability analysis in a ghost condensate background, this corresponds to
\bea
{\cal L}_{\rm Ghost\; Condensate,\, \rm quad.}^{\rm SUSY} &=& +\frac{1}{2}(\pt\p)^2 + \frac{1}{4}(\pt\p)^4 -(\pt\p)^4 (\Box \p)^2 \nn \\ && +\frac{1}{2}(\pt\chi)^2 - \frac{1}{2}(\pt\p)^2(\pt\chi)^2 - 2 (\pt\p)^4 (\pt\chi)^2 + (\pt\p \cdot\pt\chi)^2 - (\pt\p)^4 (\pt\p\cdot\pt\chi)^2 \nn \\ && +\frac{i}{2}(\psi_{,\m}\s^\m \bar\psi - \psi \s^\m \bar\psi_{,\m})\left[-1-\frac{1}{2}(\pt\p)^2\right] -\p_{\m}\p_{,\n}\frac{i}{2}(\psi^{,\n}\s^\m \bar\psi - \psi \s^\m \bar\psi^{,\n}).
\eea

\subsection{A New Form of Supersymmetry Breaking}

The supersymmetric ghost condensate is unusual in yet another way. Consider the supersymmetry transformation of the spinor,
\bea \delta \psi &=& i \sqrt{2} \sigma^\m \bar\xi \pt_\m A + \sqrt{2} \xi F\,.
\eea
Ordinarily, spontaneous breaking of supersymmetry is achieved by having a non-zero, constant vev of the dimension-two auxiliary field $F$, thus rendering the transformation inhomogeneous.  The spinor $\psi$ then becomes the Goldstone fermion of the spontaneously broken supersymmetry.

With the ghost condensate, we find ourselves in a new situation. In this vacuum, the vev of $F$ vanishes.
Now, however, supersymmetry is broken by the scalar field $A$ getting a non-zero and, moreover, linearly time-dependent vev $\langle \dot{A} \rangle = \langle \dot\p \rangle/\sqrt{2}=c/\sqrt{2}$, where we restore the arbitrary dimension-two constant. Therefore,
\bea
\d\psi &=& i \sqrt{2} \sigma^\m \bar\xi \pt_\m \delta A =i \s^0 \bar\xi c \ .
\eea
As previously, the fermion transforms inhomogeneously and, hence, supersymmetry is spontaneously broken. For the ghost condensate, however, the inhomogeneous term arises from the linear time-dependent vev of $\phi$ rather than from the $F$-term.
The scale of supersymmetry breaking corresponds to the scale of the ghost condensate.

It would be of interest to explore this mechanism within the context of supergravity. There, one would expect the Goldstone fermion to be eaten by the gravitino, and to render the latter massive. However, because of the wrong-sign spatial kinetic term of the spinor in a ghost condensate background --- as discussed in the previous subsection --- there may well be subtleties involved. We leave this intriguing question to future work.

\section{Discussion}
\label{recap}

In this paper we have established a formalism which enables us to write the supersymmetric extensions of scalar field theories of the $P(X,\phi)$ type. This is accomplished by first constructing the supersymmetric version of $X^2.$ Using this as a building block for higher powers of $X,$ any function $P(X,\phi)$ can be reconstructed as a series expansion. Immediate applications of this formalism are supersymmetric versions of DBI-type actions and of the ghost condensate model, which can be embedded in cosmological scenarios of the very early universe.

The ghost condensate theory was studied in some detail. We discovered that it leads to a new form of spontaneous supersymmetry breaking where it is not the auxiliary field but, rather, the dynamical scalar field itself that acquires a vev. This has a number of interesting properties. In particular, since the vacuum breaks Lorentz invariance, the Lagrangian describing the fluctuations of the two real scalars and the spinor belonging to a chiral multiplet is explicitly Lorentz-violating.
At first sight, the scalar degrees of freedom appear to be unstable, but we have shown how their instabilities can be cured. For the spinor, the situation is more complicated, as it appears to have a wrong-sign spatial kinetic term. This (possible) instability can be removed by extending the framework under consideration to more general higher-derivative scalar field theories, including models of the Galileon type. Such theories are closely related to $P(X,\phi)$ theories~\cite{deRham:2010eu}, and will be the subject of a following paper.

We can foresee a number of applications and extensions of our current work. For example, the DBI action is used in many models of the early universe that are derived from (or inspired by) string theory, and in many of these situations the action should be formulated in a supersymmetric way. It will be interesting to see if our results here can lead to new insights in that respect. Also, for early universe models of the ekpyrotic type, in which the current expanding phase of the universe is preceded by a contracting phase, the bounce that links these two phases must necessarily involve a violation of the NEC. The $P(X,\phi)$ theories are candidates for an effective description of this type of dynamics, and our work can be used to study this in a supersymmetric way, as may well be appropriate at higher energy.

An important, but laborious, extension of this work would be to re-formulate the higher derivative scalar actions in supergravity. Not only would such an extension constitute the most appropriate setting for discussing cosmological applications, but it would also be interesting to examine the fate of the Goldstone fermion generated in the supersymmetric ghost condensate model.
Finally, we hope that our results can help elucidate the connection of $P(X,\phi)$ theories with string theory. It is, at present, not entirely clear whether the ghost condensate model can be derived as an effective theory from string theory.
We believe our superfield expressions can be helpful in clarifying this issue, and furthering our understanding of the relationship between string compactifications and effective field theories.

\begin{acknowledgments}

We would like to thank Lam Hui, Alberto Nicolis, Paul Steinhardt and Daniel Wesley for discussions, and Daniel Baumann and Daniel Green for pointing out to us the existence of a second supersymmetric extension of $X^2$. J.L.L. expresses his thanks to the University of Pennsylvania, for its hospitality while this work was being completed. J.L.L. is supported by a Starting Grant from the European Research Council. J.K. and B.A.O. are supported in part by the DOE under contract No. DE-AC02-76-ER-03071, the NSF under grant No. 1001296, and by the Alfred P. Sloan Foundation (JK).

\end{acknowledgments}

\section*{Appendix: $P\left(\frac{1}{2}\dot\p^2 + \frac{1}{2}\dot\chi^2\right)$ and NEC violation}

When the time-dependent part of the Lagrangian is $O(2)$-symmetric under the interchange of $\p$ and $\chi$, that is, when the Lagrangian is a function of $\dot\p^2 + \dot\chi^2,$ one can prove that whenever the NEC is violated a ghost must appear, regardless of what the time-dependent vacuum might be. The implication of this is the following:  if one wants a theory with controlled and viable violations of the NEC, then the symmetry between $\p$ and $\chi$ must be broken. Here is the proof.

Assume that the purely time-dependent Lagrangian is given by
\be
{\cal L}=P(Y)\,,
\ee
where $Y \equiv \frac{1}{2} \dot\p^2 + \frac{1}{2} \dot\chi^2.$ Then the energy density and pressure are %
\bea
\rho &=& 2 Y P_{,Y} -P\;; \\
p &=& P\,,
\eea
and the condition for NEC violation is
\be
\rho + p = 2 Y P_{,Y} < 0\,.
\ee
Since $Y>0$ by definition, the condition for NEC violation becomes
 \be
 P_{,Y} < 0\,.
 \ee
 We can assess the stability properties of the theory by expanding the Lagrangian to quadratic order in the field fluctuations. We find
 \bea {\cal L}_{\rm quad.} &=& \frac{1}{2} P_{,\dot\p \dot\p}(\dot{\delta \p})^2+ P_{,\dot\p\dot\chi}\dot{\delta \p}\dot{\delta \chi} + \frac{1}{2} P_{,\dot{\chi}\dot{\chi}}(\dot{\delta \chi})^2 \nn \\ &=& \frac{1}{2}(P_{,YY}\dot\p^2+P_{,Y})(\dot{\delta\zeta})^2 + \frac{P_{,Y}[P_{,YY}(\dot\p^2+\dot\chi^2)+P_{,Y}]}{2(P_{,YY}\dot\p^2+P_{,Y})}(\dot{\delta \chi})^2,
 \eea
 where the cross term is removed by defining the new variable $\dot{\delta \zeta}\equiv \dot{\delta \p} + \frac{P_{,YY}\dot\p\dot\chi}{P_{,YY}\dot\p^2+P_{,Y}}\dot{\delta \chi}$. Hence, the conditions for stability are
\bea
 P_{,YY}\dot\p^2+P_{,Y} &>& 0\;; \\
 P_{,Y}\left[P_{,YY}(\dot\p^2+\dot\chi^2)+P_{,Y}\right] &>& 0\,.
 \eea
 These cannot be satisfied simultaneously with the NEC-violation condition $P_{,Y}<0$. This proves that there must be a ghost field as soon as the NEC is violated.


\begin{thebibliography}{66}
\expandafter\ifx\csname natexlab\endcsname\relax\def\natexlab#1{#1}\fi
\expandafter\ifx\csname bibnamefont\endcsname\relax
  \def\bibnamefont#1{#1}\fi
\expandafter\ifx\csname bibfnamefont\endcsname\relax
  \def\bibfnamefont#1{#1}\fi
\expandafter\ifx\csname citenamefont\endcsname\relax
  \def\citenamefont#1{#1}\fi
\expandafter\ifx\csname url\endcsname\relax
  \def\url#1{\texttt{#1}}\fi
\expandafter\ifx\csname urlprefix\endcsname\relax\def\urlprefix{URL }\fi
\providecommand{\bibinfo}[2]{#2}
\providecommand{\eprint}[2][]{\url{#2}}

\bibitem[{\citenamefont{Silverstein and Tong}(2004)}]{Silverstein:2003hf}
\bibinfo{author}{\bibfnamefont{E.}~\bibnamefont{Silverstein}} \bibnamefont{and}
  \bibinfo{author}{\bibfnamefont{D.}~\bibnamefont{Tong}},
  \bibinfo{journal}{Phys. Rev.} \textbf{\bibinfo{volume}{D70}},
  \bibinfo{pages}{103505} (\bibinfo{year}{2004}), \eprint{hep-th/0310221}.

\bibitem[{\citenamefont{Alishahiha et~al.}(2004)\citenamefont{Alishahiha,
  Silverstein, and Tong}}]{Alishahiha:2004eh}
\bibinfo{author}{\bibfnamefont{M.}~\bibnamefont{Alishahiha}},
  \bibinfo{author}{\bibfnamefont{E.}~\bibnamefont{Silverstein}},
  \bibnamefont{and} \bibinfo{author}{\bibfnamefont{D.}~\bibnamefont{Tong}},
  \bibinfo{journal}{Phys. Rev.} \textbf{\bibinfo{volume}{D70}},
  \bibinfo{pages}{123505} (\bibinfo{year}{2004}), \eprint{hep-th/0404084}.

\bibitem[{\citenamefont{Babich et~al.}(2004)\citenamefont{Babich, Creminelli,
  and Zaldarriaga}}]{Babich:2004gb}
\bibinfo{author}{\bibfnamefont{D.}~\bibnamefont{Babich}},
  \bibinfo{author}{\bibfnamefont{P.}~\bibnamefont{Creminelli}},
  \bibnamefont{and}
  \bibinfo{author}{\bibfnamefont{M.}~\bibnamefont{Zaldarriaga}},
  \bibinfo{journal}{JCAP} \textbf{\bibinfo{volume}{0408}}, \bibinfo{pages}{009}
  (\bibinfo{year}{2004}), \eprint{astro-ph/0405356}.

\bibitem[{\citenamefont{Khoury et~al.}(2001)\citenamefont{Khoury, Ovrut,
  Steinhardt, and Turok}}]{Khoury:2001wf}
\bibinfo{author}{\bibfnamefont{J.}~\bibnamefont{Khoury}},
  \bibinfo{author}{\bibfnamefont{B.~A.} \bibnamefont{Ovrut}},
  \bibinfo{author}{\bibfnamefont{P.~J.} \bibnamefont{Steinhardt}},
  \bibnamefont{and} \bibinfo{author}{\bibfnamefont{N.}~\bibnamefont{Turok}},
  \bibinfo{journal}{Phys. Rev.} \textbf{\bibinfo{volume}{D64}},
  \bibinfo{pages}{123522} (\bibinfo{year}{2001}), \eprint{hep-th/0103239}.

\bibitem[{\citenamefont{Donagi et~al.}(2001)\citenamefont{Donagi, Khoury,
  Ovrut, Steinhardt, and Turok}}]{Donagi:2001fs}
\bibinfo{author}{\bibfnamefont{R.~Y.} \bibnamefont{Donagi}},
  \bibinfo{author}{\bibfnamefont{J.}~\bibnamefont{Khoury}},
  \bibinfo{author}{\bibfnamefont{B.~A.} \bibnamefont{Ovrut}},
  \bibinfo{author}{\bibfnamefont{P.~J.} \bibnamefont{Steinhardt}},
  \bibnamefont{and} \bibinfo{author}{\bibfnamefont{N.}~\bibnamefont{Turok}},
  \bibinfo{journal}{JHEP} \textbf{\bibinfo{volume}{11}}, \bibinfo{pages}{041}
  (\bibinfo{year}{2001}), \eprint{hep-th/0105199}.

\bibitem[{\citenamefont{Khoury et~al.}(2002{\natexlab{a}})\citenamefont{Khoury,
  Ovrut, Seiberg, Steinhardt, and Turok}}]{Khoury:2001bz}
\bibinfo{author}{\bibfnamefont{J.}~\bibnamefont{Khoury}},
  \bibinfo{author}{\bibfnamefont{B.~A.} \bibnamefont{Ovrut}},
  \bibinfo{author}{\bibfnamefont{N.}~\bibnamefont{Seiberg}},
  \bibinfo{author}{\bibfnamefont{P.~J.} \bibnamefont{Steinhardt}},
  \bibnamefont{and} \bibinfo{author}{\bibfnamefont{N.}~\bibnamefont{Turok}},
  \bibinfo{journal}{Phys. Rev.} \textbf{\bibinfo{volume}{D65}},
  \bibinfo{pages}{086007} (\bibinfo{year}{2002}{\natexlab{a}}),
  \eprint{hep-th/0108187}.

\bibitem[{\citenamefont{Khoury et~al.}(2002{\natexlab{b}})\citenamefont{Khoury,
  Ovrut, Steinhardt, and Turok}}]{Khoury:2001zk}
\bibinfo{author}{\bibfnamefont{J.}~\bibnamefont{Khoury}},
  \bibinfo{author}{\bibfnamefont{B.~A.} \bibnamefont{Ovrut}},
  \bibinfo{author}{\bibfnamefont{P.~J.} \bibnamefont{Steinhardt}},
  \bibnamefont{and} \bibinfo{author}{\bibfnamefont{N.}~\bibnamefont{Turok}},
  \bibinfo{journal}{Phys. Rev.} \textbf{\bibinfo{volume}{D66}},
  \bibinfo{pages}{046005} (\bibinfo{year}{2002}{\natexlab{b}}),
  \eprint{hep-th/0109050}.

\bibitem[{\citenamefont{Craps and Ovrut}(2004)}]{Craps:2003ai}
\bibinfo{author}{\bibfnamefont{B.}~\bibnamefont{Craps}} \bibnamefont{and}
  \bibinfo{author}{\bibfnamefont{B.~A.} \bibnamefont{Ovrut}},
  \bibinfo{journal}{Phys. Rev.} \textbf{\bibinfo{volume}{D69}},
  \bibinfo{pages}{066001} (\bibinfo{year}{2004}), \eprint{hep-th/0308057}.

\bibitem[{\citenamefont{Khoury et~al.}(2003)\citenamefont{Khoury, Steinhardt,
  and Turok}}]{Khoury:2003vb}
\bibinfo{author}{\bibfnamefont{J.}~\bibnamefont{Khoury}},
  \bibinfo{author}{\bibfnamefont{P.~J.} \bibnamefont{Steinhardt}},
  \bibnamefont{and} \bibinfo{author}{\bibfnamefont{N.}~\bibnamefont{Turok}},
  \bibinfo{journal}{Phys. Rev. Lett.} \textbf{\bibinfo{volume}{91}},
  \bibinfo{pages}{161301} (\bibinfo{year}{2003}), \eprint{astro-ph/0302012}.

\bibitem[{\citenamefont{Khoury et~al.}(2004)\citenamefont{Khoury, Steinhardt,
  and Turok}}]{Khoury:2003rt}
\bibinfo{author}{\bibfnamefont{J.}~\bibnamefont{Khoury}},
  \bibinfo{author}{\bibfnamefont{P.~J.} \bibnamefont{Steinhardt}},
  \bibnamefont{and} \bibinfo{author}{\bibfnamefont{N.}~\bibnamefont{Turok}},
  \bibinfo{journal}{Phys. Rev. Lett.} \textbf{\bibinfo{volume}{92}},
  \bibinfo{pages}{031302} (\bibinfo{year}{2004}), \eprint{hep-th/0307132}.

\bibitem[{\citenamefont{Khoury}(2004)}]{Khoury:2004xi}
\bibinfo{author}{\bibfnamefont{J.}~\bibnamefont{Khoury}}
  (\bibinfo{year}{2004}), \eprint{astro-ph/0401579}.

\bibitem[{\citenamefont{Lehners et~al.}(2007)\citenamefont{Lehners, McFadden,
  Turok, and Steinhardt}}]{Lehners:2007ac}
\bibinfo{author}{\bibfnamefont{J.-L.} \bibnamefont{Lehners}},
  \bibinfo{author}{\bibfnamefont{P.}~\bibnamefont{McFadden}},
  \bibinfo{author}{\bibfnamefont{N.}~\bibnamefont{Turok}}, \bibnamefont{and}
  \bibinfo{author}{\bibfnamefont{P.~J.} \bibnamefont{Steinhardt}},
  \bibinfo{journal}{Phys. Rev.} \textbf{\bibinfo{volume}{D76}},
  \bibinfo{pages}{103501} (\bibinfo{year}{2007}), \eprint{hep-th/0702153}.

\bibitem[{\citenamefont{Buchbinder
  et~al.}(2007{\natexlab{a}})\citenamefont{Buchbinder, Khoury, and
  Ovrut}}]{Buchbinder:2007ad}
\bibinfo{author}{\bibfnamefont{E.~I.} \bibnamefont{Buchbinder}},
  \bibinfo{author}{\bibfnamefont{J.}~\bibnamefont{Khoury}}, \bibnamefont{and}
  \bibinfo{author}{\bibfnamefont{B.~A.} \bibnamefont{Ovrut}},
  \bibinfo{journal}{Phys. Rev.} \textbf{\bibinfo{volume}{D76}},
  \bibinfo{pages}{123503} (\bibinfo{year}{2007}{\natexlab{a}}),
  \eprint{hep-th/0702154}.

\bibitem[{\citenamefont{Buchbinder
  et~al.}(2007{\natexlab{b}})\citenamefont{Buchbinder, Khoury, and
  Ovrut}}]{Buchbinder:2007tw}
\bibinfo{author}{\bibfnamefont{E.~I.} \bibnamefont{Buchbinder}},
  \bibinfo{author}{\bibfnamefont{J.}~\bibnamefont{Khoury}}, \bibnamefont{and}
  \bibinfo{author}{\bibfnamefont{B.~A.} \bibnamefont{Ovrut}},
  \bibinfo{journal}{JHEP} \textbf{\bibinfo{volume}{11}}, \bibinfo{pages}{076}
  (\bibinfo{year}{2007}{\natexlab{b}}), \eprint{0706.3903}.

\bibitem[{\citenamefont{Creminelli and Senatore}(2007)}]{Creminelli:2007aq}
\bibinfo{author}{\bibfnamefont{P.}~\bibnamefont{Creminelli}} \bibnamefont{and}
  \bibinfo{author}{\bibfnamefont{L.}~\bibnamefont{Senatore}},
  \bibinfo{journal}{JCAP} \textbf{\bibinfo{volume}{0711}}, \bibinfo{pages}{010}
  (\bibinfo{year}{2007}), \eprint{hep-th/0702165}.

\bibitem[{\citenamefont{Koyama et~al.}(2007{\natexlab{a}})\citenamefont{Koyama,
  Mizuno, and Wands}}]{Koyama:2007ag}
\bibinfo{author}{\bibfnamefont{K.}~\bibnamefont{Koyama}},
  \bibinfo{author}{\bibfnamefont{S.}~\bibnamefont{Mizuno}}, \bibnamefont{and}
  \bibinfo{author}{\bibfnamefont{D.}~\bibnamefont{Wands}},
  \bibinfo{journal}{Class. Quant. Grav.} \textbf{\bibinfo{volume}{24}},
  \bibinfo{pages}{3919} (\bibinfo{year}{2007}{\natexlab{a}}),
  \eprint{0704.1152}.

\bibitem[{\citenamefont{Koyama and Wands}(2007)}]{Koyama:2007mg}
\bibinfo{author}{\bibfnamefont{K.}~\bibnamefont{Koyama}} \bibnamefont{and}
  \bibinfo{author}{\bibfnamefont{D.}~\bibnamefont{Wands}},
  \bibinfo{journal}{JCAP} \textbf{\bibinfo{volume}{0704}}, \bibinfo{pages}{008}
  (\bibinfo{year}{2007}), \eprint{hep-th/0703040}.

\bibitem[{\citenamefont{Gasperini and Veneziano}(1993)}]{Gasperini:1992em}
\bibinfo{author}{\bibfnamefont{M.}~\bibnamefont{Gasperini}} \bibnamefont{and}
  \bibinfo{author}{\bibfnamefont{G.}~\bibnamefont{Veneziano}},
  \bibinfo{journal}{Astropart. Phys.} \textbf{\bibinfo{volume}{1}},
  \bibinfo{pages}{317} (\bibinfo{year}{1993}), \eprint{hep-th/9211021}.

\bibitem[{\citenamefont{Gasperini and Veneziano}(2003)}]{Gasperini:2002bn}
\bibinfo{author}{\bibfnamefont{M.}~\bibnamefont{Gasperini}} \bibnamefont{and}
  \bibinfo{author}{\bibfnamefont{G.}~\bibnamefont{Veneziano}},
  \bibinfo{journal}{Phys. Rept.} \textbf{\bibinfo{volume}{373}},
  \bibinfo{pages}{1} (\bibinfo{year}{2003}), \eprint{hep-th/0207130}.

\bibitem[{\citenamefont{Finelli and Brandenberger}(2002)}]{Finelli:2001sr}
\bibinfo{author}{\bibfnamefont{F.}~\bibnamefont{Finelli}} \bibnamefont{and}
  \bibinfo{author}{\bibfnamefont{R.}~\bibnamefont{Brandenberger}},
  \bibinfo{journal}{Phys. Rev.} \textbf{\bibinfo{volume}{D65}},
  \bibinfo{pages}{103522} (\bibinfo{year}{2002}), \eprint{hep-th/0112249}.

\bibitem[{\citenamefont{Creminelli et~al.}(2010)\citenamefont{Creminelli,
  Nicolis, and Trincherini}}]{Creminelli:2010ba}
\bibinfo{author}{\bibfnamefont{P.}~\bibnamefont{Creminelli}},
  \bibinfo{author}{\bibfnamefont{A.}~\bibnamefont{Nicolis}}, \bibnamefont{and}
  \bibinfo{author}{\bibfnamefont{E.}~\bibnamefont{Trincherini}},
  \bibinfo{journal}{JCAP} \textbf{\bibinfo{volume}{1011}}, \bibinfo{pages}{021}
  (\bibinfo{year}{2010}), \eprint{1007.0027}.

\bibitem[{\citenamefont{Lehners}(2008)}]{Lehners:2008vx}
\bibinfo{author}{\bibfnamefont{J.-L.} \bibnamefont{Lehners}},
  \bibinfo{journal}{Phys. Rept.} \textbf{\bibinfo{volume}{465}},
  \bibinfo{pages}{223} (\bibinfo{year}{2008}), \eprint{0806.1245}.

\bibitem[{\citenamefont{Lukas et~al.}(1998{\natexlab{a}})\citenamefont{Lukas,
  Ovrut, and Waldram}}]{Lukas:1997fg}
\bibinfo{author}{\bibfnamefont{A.}~\bibnamefont{Lukas}},
  \bibinfo{author}{\bibfnamefont{B.~A.} \bibnamefont{Ovrut}}, \bibnamefont{and}
  \bibinfo{author}{\bibfnamefont{D.}~\bibnamefont{Waldram}},
  \bibinfo{journal}{Nucl. Phys.} \textbf{\bibinfo{volume}{B532}},
  \bibinfo{pages}{43} (\bibinfo{year}{1998}{\natexlab{a}}),
  \eprint{hep-th/9710208}.

\bibitem[{\citenamefont{Lukas et~al.}(1999{\natexlab{a}})\citenamefont{Lukas,
  Ovrut, Stelle, and Waldram}}]{Lukas:1998yy}
\bibinfo{author}{\bibfnamefont{A.}~\bibnamefont{Lukas}},
  \bibinfo{author}{\bibfnamefont{B.~A.} \bibnamefont{Ovrut}},
  \bibinfo{author}{\bibfnamefont{K.~S.} \bibnamefont{Stelle}},
  \bibnamefont{and} \bibinfo{author}{\bibfnamefont{D.}~\bibnamefont{Waldram}},
  \bibinfo{journal}{Phys. Rev.} \textbf{\bibinfo{volume}{D59}},
  \bibinfo{pages}{086001} (\bibinfo{year}{1999}{\natexlab{a}}),
  \eprint{hep-th/9803235}.

\bibitem[{\citenamefont{Lukas et~al.}(1999{\natexlab{b}})\citenamefont{Lukas,
  Ovrut, Stelle, and Waldram}}]{Lukas:1998tt}
\bibinfo{author}{\bibfnamefont{A.}~\bibnamefont{Lukas}},
  \bibinfo{author}{\bibfnamefont{B.~A.} \bibnamefont{Ovrut}},
  \bibinfo{author}{\bibfnamefont{K.~S.} \bibnamefont{Stelle}},
  \bibnamefont{and} \bibinfo{author}{\bibfnamefont{D.}~\bibnamefont{Waldram}},
  \bibinfo{journal}{Nucl. Phys.} \textbf{\bibinfo{volume}{B552}},
  \bibinfo{pages}{246} (\bibinfo{year}{1999}{\natexlab{b}}),
  \eprint{hep-th/9806051}.

\bibitem[{\citenamefont{Lukas et~al.}(1999{\natexlab{c}})\citenamefont{Lukas,
  Ovrut, and Waldram}}]{Lukas:1998qs}
\bibinfo{author}{\bibfnamefont{A.}~\bibnamefont{Lukas}},
  \bibinfo{author}{\bibfnamefont{B.~A.} \bibnamefont{Ovrut}}, \bibnamefont{and}
  \bibinfo{author}{\bibfnamefont{D.}~\bibnamefont{Waldram}},
  \bibinfo{journal}{Phys. Rev.} \textbf{\bibinfo{volume}{D60}},
  \bibinfo{pages}{086001} (\bibinfo{year}{1999}{\natexlab{c}}),
  \eprint{hep-th/9806022}.

\bibitem[{\citenamefont{Lukas et~al.}(1999{\natexlab{d}})\citenamefont{Lukas,
  Ovrut, and Waldram}}]{Lukas:1998ew}
\bibinfo{author}{\bibfnamefont{A.}~\bibnamefont{Lukas}},
  \bibinfo{author}{\bibfnamefont{B.~A.} \bibnamefont{Ovrut}}, \bibnamefont{and}
  \bibinfo{author}{\bibfnamefont{D.}~\bibnamefont{Waldram}},
  \bibinfo{journal}{Nucl. Phys.} \textbf{\bibinfo{volume}{B540}},
  \bibinfo{pages}{230} (\bibinfo{year}{1999}{\natexlab{d}}),
  \eprint{hep-th/9801087}.

\bibitem[{\citenamefont{Lukas et~al.}(2000)\citenamefont{Lukas, Ovrut, and
  Waldram}}]{Lukas:1999yn}
\bibinfo{author}{\bibfnamefont{A.}~\bibnamefont{Lukas}},
  \bibinfo{author}{\bibfnamefont{B.~A.} \bibnamefont{Ovrut}}, \bibnamefont{and}
  \bibinfo{author}{\bibfnamefont{D.}~\bibnamefont{Waldram}},
  \bibinfo{journal}{Phys. Rev.} \textbf{\bibinfo{volume}{D61}},
  \bibinfo{pages}{023506} (\bibinfo{year}{2000}), \eprint{hep-th/9902071}.

\bibitem[{\citenamefont{Braun et~al.}(2006)\citenamefont{Braun, He, Ovrut, and
  Pantev}}]{Braun:2005nv}
\bibinfo{author}{\bibfnamefont{V.}~\bibnamefont{Braun}},
  \bibinfo{author}{\bibfnamefont{Y.-H.} \bibnamefont{He}},
  \bibinfo{author}{\bibfnamefont{B.~A.} \bibnamefont{Ovrut}}, \bibnamefont{and}
  \bibinfo{author}{\bibfnamefont{T.}~\bibnamefont{Pantev}},
  \bibinfo{journal}{JHEP} \textbf{\bibinfo{volume}{05}}, \bibinfo{pages}{043}
  (\bibinfo{year}{2006}), \eprint{hep-th/0512177}.

\bibitem[{\citenamefont{Braun et~al.}(2005)\citenamefont{Braun, He, Ovrut, and
  Pantev}}]{Braun:2005ux}
\bibinfo{author}{\bibfnamefont{V.}~\bibnamefont{Braun}},
  \bibinfo{author}{\bibfnamefont{Y.-H.} \bibnamefont{He}},
  \bibinfo{author}{\bibfnamefont{B.~A.} \bibnamefont{Ovrut}}, \bibnamefont{and}
  \bibinfo{author}{\bibfnamefont{T.}~\bibnamefont{Pantev}},
  \bibinfo{journal}{Phys. Lett.} \textbf{\bibinfo{volume}{B618}},
  \bibinfo{pages}{252} (\bibinfo{year}{2005}), \eprint{hep-th/0501070}.

\bibitem[{\citenamefont{Lukas et~al.}(1998{\natexlab{b}})\citenamefont{Lukas,
  Ovrut, and Waldram}}]{Lukas:1996zq}
\bibinfo{author}{\bibfnamefont{A.}~\bibnamefont{Lukas}},
  \bibinfo{author}{\bibfnamefont{B.~A.} \bibnamefont{Ovrut}}, \bibnamefont{and}
  \bibinfo{author}{\bibfnamefont{D.}~\bibnamefont{Waldram}},
  \bibinfo{journal}{Nucl. Phys.} \textbf{\bibinfo{volume}{B509}},
  \bibinfo{pages}{169} (\bibinfo{year}{1998}{\natexlab{b}}),
  \eprint{hep-th/9611204}.

\bibitem[{\citenamefont{Lukas et~al.}(1997{\natexlab{a}})\citenamefont{Lukas,
  Ovrut, and Waldram}}]{Lukas:1996ee}
\bibinfo{author}{\bibfnamefont{A.}~\bibnamefont{Lukas}},
  \bibinfo{author}{\bibfnamefont{B.~A.} \bibnamefont{Ovrut}}, \bibnamefont{and}
  \bibinfo{author}{\bibfnamefont{D.}~\bibnamefont{Waldram}},
  \bibinfo{journal}{Phys. Lett.} \textbf{\bibinfo{volume}{B393}},
  \bibinfo{pages}{65} (\bibinfo{year}{1997}{\natexlab{a}}),
  \eprint{hep-th/9608195}.

\bibitem[{\citenamefont{Lukas et~al.}(1997{\natexlab{b}})\citenamefont{Lukas,
  Ovrut, and Waldram}}]{Lukas:1996iq}
\bibinfo{author}{\bibfnamefont{A.}~\bibnamefont{Lukas}},
  \bibinfo{author}{\bibfnamefont{B.~A.} \bibnamefont{Ovrut}}, \bibnamefont{and}
  \bibinfo{author}{\bibfnamefont{D.}~\bibnamefont{Waldram}},
  \bibinfo{journal}{Nucl. Phys.} \textbf{\bibinfo{volume}{B495}},
  \bibinfo{pages}{365} (\bibinfo{year}{1997}{\natexlab{b}}),
  \eprint{hep-th/9610238}.

\bibitem[{\citenamefont{Lukas and Ovrut}(1998)}]{Lukas:1997yc}
\bibinfo{author}{\bibfnamefont{A.}~\bibnamefont{Lukas}} \bibnamefont{and}
  \bibinfo{author}{\bibfnamefont{B.~A.} \bibnamefont{Ovrut}},
  \bibinfo{journal}{Phys. Lett.} \textbf{\bibinfo{volume}{B437}},
  \bibinfo{pages}{291} (\bibinfo{year}{1998}), \eprint{hep-th/9709030}.

\bibitem[{\citenamefont{Brandle et~al.}(2001)\citenamefont{Brandle, Lukas, and
  Ovrut}}]{Brandle:2000qp}
\bibinfo{author}{\bibfnamefont{M.}~\bibnamefont{Brandle}},
  \bibinfo{author}{\bibfnamefont{A.}~\bibnamefont{Lukas}}, \bibnamefont{and}
  \bibinfo{author}{\bibfnamefont{B.~A.} \bibnamefont{Ovrut}},
  \bibinfo{journal}{Phys. Rev.} \textbf{\bibinfo{volume}{D63}},
  \bibinfo{pages}{026003} (\bibinfo{year}{2001}), \eprint{hep-th/0003256}.

\bibitem[{\citenamefont{Buchbinder et~al.}(2003)\citenamefont{Buchbinder,
  Donagi, and Ovrut}}]{Buchbinder:2002ic}
\bibinfo{author}{\bibfnamefont{E.~I.} \bibnamefont{Buchbinder}},
  \bibinfo{author}{\bibfnamefont{R.}~\bibnamefont{Donagi}}, \bibnamefont{and}
  \bibinfo{author}{\bibfnamefont{B.~A.} \bibnamefont{Ovrut}},
  \bibinfo{journal}{Nucl. Phys.} \textbf{\bibinfo{volume}{B653}},
  \bibinfo{pages}{400} (\bibinfo{year}{2003}), \eprint{hep-th/0205190}.

\bibitem[{\citenamefont{Buchbinder et~al.}(2002)\citenamefont{Buchbinder,
  Donagi, and Ovrut}}]{Buchbinder:2002ji}
\bibinfo{author}{\bibfnamefont{E.}~\bibnamefont{Buchbinder}},
  \bibinfo{author}{\bibfnamefont{R.}~\bibnamefont{Donagi}}, \bibnamefont{and}
  \bibinfo{author}{\bibfnamefont{B.~A.} \bibnamefont{Ovrut}},
  \bibinfo{journal}{JHEP} \textbf{\bibinfo{volume}{06}}, \bibinfo{pages}{054}
  (\bibinfo{year}{2002}), \eprint{hep-th/0202084}.

\bibitem[{\citenamefont{Lima et~al.}(2001)\citenamefont{Lima, Ovrut, Park, and
  Reinbacher}}]{Lima:2001jc}
\bibinfo{author}{\bibfnamefont{E.}~\bibnamefont{Lima}},
  \bibinfo{author}{\bibfnamefont{B.~A.} \bibnamefont{Ovrut}},
  \bibinfo{author}{\bibfnamefont{J.}~\bibnamefont{Park}}, \bibnamefont{and}
  \bibinfo{author}{\bibfnamefont{R.}~\bibnamefont{Reinbacher}},
  \bibinfo{journal}{Nucl. Phys.} \textbf{\bibinfo{volume}{B614}},
  \bibinfo{pages}{117} (\bibinfo{year}{2001}), \eprint{hep-th/0101049}.

\bibitem[{\citenamefont{Lima et~al.}(2002)\citenamefont{Lima, Ovrut, and
  Park}}]{Lima:2001nh}
\bibinfo{author}{\bibfnamefont{E.}~\bibnamefont{Lima}},
  \bibinfo{author}{\bibfnamefont{B.~A.} \bibnamefont{Ovrut}}, \bibnamefont{and}
  \bibinfo{author}{\bibfnamefont{J.}~\bibnamefont{Park}},
  \bibinfo{journal}{Nucl. Phys.} \textbf{\bibinfo{volume}{B626}},
  \bibinfo{pages}{113} (\bibinfo{year}{2002}), \eprint{hep-th/0102046}.

\bibitem[{\citenamefont{Donagi et~al.}(1999)\citenamefont{Donagi, Ovrut, and
  Waldram}}]{Donagi:1999jp}
\bibinfo{author}{\bibfnamefont{R.}~\bibnamefont{Donagi}},
  \bibinfo{author}{\bibfnamefont{B.~A.} \bibnamefont{Ovrut}}, \bibnamefont{and}
  \bibinfo{author}{\bibfnamefont{D.}~\bibnamefont{Waldram}},
  \bibinfo{journal}{JHEP} \textbf{\bibinfo{volume}{11}}, \bibinfo{pages}{030}
  (\bibinfo{year}{1999}), \eprint{hep-th/9904054}.

\bibitem[{\citenamefont{Ovrut et~al.}(2000)\citenamefont{Ovrut, Pantev, and
  Park}}]{Ovrut:2000qi}
\bibinfo{author}{\bibfnamefont{B.~A.} \bibnamefont{Ovrut}},
  \bibinfo{author}{\bibfnamefont{T.}~\bibnamefont{Pantev}}, \bibnamefont{and}
  \bibinfo{author}{\bibfnamefont{J.}~\bibnamefont{Park}},
  \bibinfo{journal}{JHEP} \textbf{\bibinfo{volume}{05}}, \bibinfo{pages}{045}
  (\bibinfo{year}{2000}), \eprint{hep-th/0001133}.

\bibitem[{\citenamefont{Buchbinder et~al.}(2008)\citenamefont{Buchbinder,
  Khoury, and Ovrut}}]{Buchbinder:2007at}
\bibinfo{author}{\bibfnamefont{E.~I.} \bibnamefont{Buchbinder}},
  \bibinfo{author}{\bibfnamefont{J.}~\bibnamefont{Khoury}}, \bibnamefont{and}
  \bibinfo{author}{\bibfnamefont{B.~A.} \bibnamefont{Ovrut}},
  \bibinfo{journal}{Phys. Rev. Lett.} \textbf{\bibinfo{volume}{100}},
  \bibinfo{pages}{171302} (\bibinfo{year}{2008}), \eprint{0710.5172}.

\bibitem[{\citenamefont{Koyama et~al.}(2007{\natexlab{b}})\citenamefont{Koyama,
  Mizuno, Vernizzi, and Wands}}]{Koyama:2007if}
\bibinfo{author}{\bibfnamefont{K.}~\bibnamefont{Koyama}},
  \bibinfo{author}{\bibfnamefont{S.}~\bibnamefont{Mizuno}},
  \bibinfo{author}{\bibfnamefont{F.}~\bibnamefont{Vernizzi}}, \bibnamefont{and}
  \bibinfo{author}{\bibfnamefont{D.}~\bibnamefont{Wands}},
  \bibinfo{journal}{JCAP} \textbf{\bibinfo{volume}{0711}}, \bibinfo{pages}{024}
  (\bibinfo{year}{2007}{\natexlab{b}}), \eprint{0708.4321}.

\bibitem[{\citenamefont{Lehners and
  Steinhardt}(2008{\natexlab{a}})}]{Lehners:2007wc}
\bibinfo{author}{\bibfnamefont{J.-L.} \bibnamefont{Lehners}} \bibnamefont{and}
  \bibinfo{author}{\bibfnamefont{P.~J.} \bibnamefont{Steinhardt}},
  \bibinfo{journal}{Phys. Rev.} \textbf{\bibinfo{volume}{D77}},
  \bibinfo{pages}{063533} (\bibinfo{year}{2008}{\natexlab{a}}),
  \eprint{0712.3779}.

\bibitem[{\citenamefont{Lehners and
  Steinhardt}(2008{\natexlab{b}})}]{Lehners:2008my}
\bibinfo{author}{\bibfnamefont{J.-L.} \bibnamefont{Lehners}} \bibnamefont{and}
  \bibinfo{author}{\bibfnamefont{P.~J.} \bibnamefont{Steinhardt}},
  \bibinfo{journal}{Phys. Rev.} \textbf{\bibinfo{volume}{D78}},
  \bibinfo{pages}{023506} (\bibinfo{year}{2008}{\natexlab{b}}),
  \eprint{0804.1293}.

\bibitem[{\citenamefont{Mizuno et~al.}(2008)\citenamefont{Mizuno, Koyama,
  Vernizzi, and Wands}}]{Mizuno:2008zza}
\bibinfo{author}{\bibfnamefont{S.}~\bibnamefont{Mizuno}},
  \bibinfo{author}{\bibfnamefont{K.}~\bibnamefont{Koyama}},
  \bibinfo{author}{\bibfnamefont{F.}~\bibnamefont{Vernizzi}}, \bibnamefont{and}
  \bibinfo{author}{\bibfnamefont{D.}~\bibnamefont{Wands}},
  \bibinfo{journal}{AIP Conf. Proc.} \textbf{\bibinfo{volume}{1040}},
  \bibinfo{pages}{121} (\bibinfo{year}{2008}).

\bibitem[{\citenamefont{Lehners and Steinhardt}(2009)}]{Lehners:2009qu}
\bibinfo{author}{\bibfnamefont{J.-L.} \bibnamefont{Lehners}} \bibnamefont{and}
  \bibinfo{author}{\bibfnamefont{P.~J.} \bibnamefont{Steinhardt}},
  \bibinfo{journal}{Phys. Rev.} \textbf{\bibinfo{volume}{D80}},
  \bibinfo{pages}{103520} (\bibinfo{year}{2009}), \eprint{0909.2558}.

\bibitem[{\citenamefont{Lehners and Renaux-Petel}(2009)}]{Lehners:2009ja}
\bibinfo{author}{\bibfnamefont{J.-L.} \bibnamefont{Lehners}} \bibnamefont{and}
  \bibinfo{author}{\bibfnamefont{S.}~\bibnamefont{Renaux-Petel}},
  \bibinfo{journal}{Phys. Rev.} \textbf{\bibinfo{volume}{D80}},
  \bibinfo{pages}{063503} (\bibinfo{year}{2009}), \eprint{0906.0530}.

\bibitem[{\citenamefont{Khoury and Steinhardt}(2010)}]{Khoury:2009my}
\bibinfo{author}{\bibfnamefont{J.}~\bibnamefont{Khoury}} \bibnamefont{and}
  \bibinfo{author}{\bibfnamefont{P.~J.} \bibnamefont{Steinhardt}},
  \bibinfo{journal}{Phys. Rev. Lett.} \textbf{\bibinfo{volume}{104}},
  \bibinfo{pages}{091301} (\bibinfo{year}{2010}), \eprint{0910.2230}.

\bibitem[{\citenamefont{Khoury and Miller}(2010)}]{Khoury:2010gw}
\bibinfo{author}{\bibfnamefont{J.}~\bibnamefont{Khoury}} \bibnamefont{and}
  \bibinfo{author}{\bibfnamefont{G.~E.~J.} \bibnamefont{Miller}}
  (\bibinfo{year}{2010}), \eprint{1012.0846}.

\bibitem[{\citenamefont{Lehners}(2010)}]{Lehners:2010fy}
\bibinfo{author}{\bibfnamefont{J.-L.} \bibnamefont{Lehners}},
  \bibinfo{journal}{Adv. Astron.} \textbf{\bibinfo{volume}{2010}},
  \bibinfo{pages}{903907} (\bibinfo{year}{2010}), \eprint{1001.3125}.

\bibitem[{\citenamefont{Boyle et~al.}(2004)\citenamefont{Boyle, Steinhardt, and
  Turok}}]{Boyle:2003km}
\bibinfo{author}{\bibfnamefont{L.~A.} \bibnamefont{Boyle}},
  \bibinfo{author}{\bibfnamefont{P.~J.} \bibnamefont{Steinhardt}},
  \bibnamefont{and} \bibinfo{author}{\bibfnamefont{N.}~\bibnamefont{Turok}},
  \bibinfo{journal}{Phys. Rev.} \textbf{\bibinfo{volume}{D69}},
  \bibinfo{pages}{127302} (\bibinfo{year}{2004}), \eprint{hep-th/0307170}.

\bibitem[{\citenamefont{Baumann et~al.}(2007)\citenamefont{Baumann, Steinhardt,
  Takahashi, and Ichiki}}]{Baumann:2007zm}
\bibinfo{author}{\bibfnamefont{D.}~\bibnamefont{Baumann}},
  \bibinfo{author}{\bibfnamefont{P.~J.} \bibnamefont{Steinhardt}},
  \bibinfo{author}{\bibfnamefont{K.}~\bibnamefont{Takahashi}},
  \bibnamefont{and} \bibinfo{author}{\bibfnamefont{K.}~\bibnamefont{Ichiki}},
  \bibinfo{journal}{Phys. Rev.} \textbf{\bibinfo{volume}{D76}},
  \bibinfo{pages}{084019} (\bibinfo{year}{2007}), \eprint{hep-th/0703290}.

\bibitem[{\citenamefont{Dubovsky et~al.}(2006)\citenamefont{Dubovsky, Gregoire,
  Nicolis, and Rattazzi}}]{Dubovsky:2005xd}
\bibinfo{author}{\bibfnamefont{S.}~\bibnamefont{Dubovsky}},
  \bibinfo{author}{\bibfnamefont{T.}~\bibnamefont{Gregoire}},
  \bibinfo{author}{\bibfnamefont{A.}~\bibnamefont{Nicolis}}, \bibnamefont{and}
  \bibinfo{author}{\bibfnamefont{R.}~\bibnamefont{Rattazzi}},
  \bibinfo{journal}{JHEP} \textbf{\bibinfo{volume}{03}}, \bibinfo{pages}{025}
  (\bibinfo{year}{2006}), \eprint{hep-th/0512260}.

\bibitem[{\citenamefont{Creminelli et~al.}(2006)\citenamefont{Creminelli, Luty,
  Nicolis, and Senatore}}]{Creminelli:2006xe}
\bibinfo{author}{\bibfnamefont{P.}~\bibnamefont{Creminelli}},
  \bibinfo{author}{\bibfnamefont{M.~A.} \bibnamefont{Luty}},
  \bibinfo{author}{\bibfnamefont{A.}~\bibnamefont{Nicolis}}, \bibnamefont{and}
  \bibinfo{author}{\bibfnamefont{L.}~\bibnamefont{Senatore}},
  \bibinfo{journal}{JHEP} \textbf{\bibinfo{volume}{12}}, \bibinfo{pages}{080}
  (\bibinfo{year}{2006}), \eprint{hep-th/0606090}.

\bibitem[{\citenamefont{Nicolis et~al.}(2010)\citenamefont{Nicolis, Rattazzi,
  and Trincherini}}]{Nicolis:2009qm}
\bibinfo{author}{\bibfnamefont{A.}~\bibnamefont{Nicolis}},
  \bibinfo{author}{\bibfnamefont{R.}~\bibnamefont{Rattazzi}}, \bibnamefont{and}
  \bibinfo{author}{\bibfnamefont{E.}~\bibnamefont{Trincherini}},
  \bibinfo{journal}{JHEP} \textbf{\bibinfo{volume}{05}}, \bibinfo{pages}{095}
  (\bibinfo{year}{2010}), \eprint{0912.4258}.

\bibitem[{\citenamefont{Arkani-Hamed et~al.}(2004)\citenamefont{Arkani-Hamed,
  Cheng, Luty, and Mukohyama}}]{ArkaniHamed:2003uy}
\bibinfo{author}{\bibfnamefont{N.}~\bibnamefont{Arkani-Hamed}},
  \bibinfo{author}{\bibfnamefont{H.-C.} \bibnamefont{Cheng}},
  \bibinfo{author}{\bibfnamefont{M.~A.} \bibnamefont{Luty}}, \bibnamefont{and}
  \bibinfo{author}{\bibfnamefont{S.}~\bibnamefont{Mukohyama}},
  \bibinfo{journal}{JHEP} \textbf{\bibinfo{volume}{05}}, \bibinfo{pages}{074}
  (\bibinfo{year}{2004}), \eprint{hep-th/0312099}.

\bibitem[{\citenamefont{Arkani-Hamed et~al.}(2007)\citenamefont{Arkani-Hamed,
  Cheng, Luty, Mukohyama, and Wiseman}}]{ArkaniHamed:2005gu}
\bibinfo{author}{\bibfnamefont{N.}~\bibnamefont{Arkani-Hamed}},
  \bibinfo{author}{\bibfnamefont{H.-C.} \bibnamefont{Cheng}},
  \bibinfo{author}{\bibfnamefont{M.~A.} \bibnamefont{Luty}},
  \bibinfo{author}{\bibfnamefont{S.}~\bibnamefont{Mukohyama}},
  \bibnamefont{and} \bibinfo{author}{\bibfnamefont{T.}~\bibnamefont{Wiseman}},
  \bibinfo{journal}{JHEP} \textbf{\bibinfo{volume}{01}}, \bibinfo{pages}{036}
  (\bibinfo{year}{2007}), \eprint{hep-ph/0507120}.

\bibitem[{\citenamefont{Nicolis et~al.}(2009)\citenamefont{Nicolis, Rattazzi,
  and Trincherini}}]{Nicolis:2008in}
\bibinfo{author}{\bibfnamefont{A.}~\bibnamefont{Nicolis}},
  \bibinfo{author}{\bibfnamefont{R.}~\bibnamefont{Rattazzi}}, \bibnamefont{and}
  \bibinfo{author}{\bibfnamefont{E.}~\bibnamefont{Trincherini}},
  \bibinfo{journal}{Phys. Rev.} \textbf{\bibinfo{volume}{D79}},
  \bibinfo{pages}{064036} (\bibinfo{year}{2009}), \eprint{0811.2197}.

\bibitem[{\citenamefont{Dubovsky and Sibiryakov}(2006)}]{Dubovsky:2006vk}
\bibinfo{author}{\bibfnamefont{S.~L.} \bibnamefont{Dubovsky}} \bibnamefont{and}
  \bibinfo{author}{\bibfnamefont{S.~M.} \bibnamefont{Sibiryakov}},
  \bibinfo{journal}{Phys. Lett.} \textbf{\bibinfo{volume}{B638}},
  \bibinfo{pages}{509} (\bibinfo{year}{2006}), \eprint{hep-th/0603158}.

\bibitem[{\citenamefont{Adams et~al.}(2006)\citenamefont{Adams, Arkani-Hamed,
  Dubovsky, Nicolis, and Rattazzi}}]{Adams:2006sv}
\bibinfo{author}{\bibfnamefont{A.}~\bibnamefont{Adams}},
  \bibinfo{author}{\bibfnamefont{N.}~\bibnamefont{Arkani-Hamed}},
  \bibinfo{author}{\bibfnamefont{S.}~\bibnamefont{Dubovsky}},
  \bibinfo{author}{\bibfnamefont{A.}~\bibnamefont{Nicolis}}, \bibnamefont{and}
  \bibinfo{author}{\bibfnamefont{R.}~\bibnamefont{Rattazzi}},
  \bibinfo{journal}{JHEP} \textbf{\bibinfo{volume}{10}}, \bibinfo{pages}{014}
  (\bibinfo{year}{2006}), \eprint{hep-th/0602178}.

\bibitem{Aganagic:1996pe}
  M.~Aganagic, C.~Popescu and J.~H.~Schwarz,
  Phys.\ Lett.\  B {\bf 393}, 311 (1997)
  [arXiv:hep-th/9610249].

\bibitem{Aganagic:1996nn}
  M.~Aganagic, C.~Popescu and J.~H.~Schwarz,
  Nucl.\ Phys.\  B {\bf 495}, 99 (1997)
  [arXiv:hep-th/9612080].

\bibitem{Cederwall:1996ri}
  M.~Cederwall, A.~von Gussich, B.~E.~W.~Nilsson, P.~Sundell and A.~Westerberg,
  Nucl.\ Phys.\  B {\bf 490}, 179 (1997)
  [arXiv:hep-th/9611159].

\bibitem{Bergshoeff:1996tu}
  E.~Bergshoeff and P.~K.~Townsend,
  Nucl.\ Phys.\  B {\bf 490}, 145 (1997)
  [arXiv:hep-th/9611173].

\bibitem{Bazeia:2009db}
  D.~Bazeia, R.~Menezes and A.~Y.~Petrov,
  Phys.\ Lett.\  B {\bf 683}, 335 (2010)
  [arXiv:0910.2827 [hep-th]].



\bibitem[{\citenamefont{Wess and Bagger}(1992)}]{Wess:1992cp}
\bibinfo{author}{\bibfnamefont{J.}~\bibnamefont{Wess}} \bibnamefont{and}
  \bibinfo{author}{\bibfnamefont{J.}~\bibnamefont{Bagger}}
  (\bibinfo{year}{1992}), \bibinfo{note}{Princeton, USA: Univ. Pr. 259 p}.

\bibitem[{\citenamefont{Baumann and Green}(2011)}]{BaumannGreen}
\bibinfo{author}{\bibfnamefont{D.}~\bibnamefont{Baumann}} \bibnamefont{and}
  \bibinfo{author}{\bibfnamefont{D.}~\bibnamefont{Green}}
  (\bibinfo{year}{2011}), \bibinfo{note}{Private Communication}.

\bibitem[{\citenamefont{Wesley}(2008)}]{Wesley:2008de}
\bibinfo{author}{\bibfnamefont{D.~H.} \bibnamefont{Wesley}}
  (\bibinfo{year}{2008}), \eprint{0802.2106}.

\bibitem[{\citenamefont{Wesley}(2009)}]{Wesley:2008fg}
\bibinfo{author}{\bibfnamefont{D.~H.} \bibnamefont{Wesley}},
  \bibinfo{journal}{JCAP} \textbf{\bibinfo{volume}{0901}}, \bibinfo{pages}{041}
  (\bibinfo{year}{2009}), \eprint{0802.3214}.

\bibitem[{\citenamefont{Steinhardt and Wesley}(2009)}]{Steinhardt:2008nk}
\bibinfo{author}{\bibfnamefont{P.~J.} \bibnamefont{Steinhardt}}
  \bibnamefont{and} \bibinfo{author}{\bibfnamefont{D.}~\bibnamefont{Wesley}},
  \bibinfo{journal}{Phys. Rev.} \textbf{\bibinfo{volume}{D79}},
  \bibinfo{pages}{104026} (\bibinfo{year}{2009}), \eprint{0811.1614}.

\bibitem[{\citenamefont{de~Rham and Tolley}(2010)}]{deRham:2010eu}
\bibinfo{author}{\bibfnamefont{C.}~\bibnamefont{de~Rham}} \bibnamefont{and}
  \bibinfo{author}{\bibfnamefont{A.~J.} \bibnamefont{Tolley}},
  \bibinfo{journal}{JCAP} \textbf{\bibinfo{volume}{1005}}, \bibinfo{pages}{015}
  (\bibinfo{year}{2010}), \eprint{1003.5917}.

\end{thebibliography}
\end{document}